\documentclass[pre,epsfig,aps]{revtex4}
\usepackage{epsfig}
\usepackage{fancyheadings}
\usepackage{amstext,amsmath,amssymb}

\begin{document}
\title{Introduction to Phase Transitions in Random Optimization Problems}
\author{R. Monasson}
\address{Laboratoire de Physique Th\'eorique de l'ENS\\ 24 rue
  Lhomond, 75005 Paris, France}
\begin{abstract}
Notes of the lectures delivered in Les Houches during
the Summer School on Complex Systems (July 2006). 
\end{abstract}
\maketitle

\section{Introduction}

\subsection{Preamble}

The connection between the statistical physics of disordered systems
 and optimization problems
in computer science dates back from twenty years at least
\cite{Me87}. After all zero temperature statistical physics is
simply the search for the state with minimal energy, while
the main problem in combinatorial optimization is to look for the
configurations of parameters minimizing some cost function (the length
of a tour in the traveling salesman problem (TSP), the number of violated
constraints in constrained satisfaction problems, ...) \cite{Pa83}.
Yet, despite the beautiful studies of the average
 properties of the TSP, Graph partitioning, Matching, ..., based on 
the recently developed  mean-field spin glass theory \cite{Me87},  
a methodological gap between the fields could not be bridged\cite{Fu85}. 
In statistical physics statements are usually made on the 
properties of samples given some quenched disorder
distribution  such as the typical number of solutions, minimal
energy ...
In optimization, however, one is interested in solving one (or
several) particular instances of a problem, and needs efficient
ways to do so, that is, requiring a computational effort growing
not too quickly with the number of data defining the instance.
Knowing precisely the typical properties for a given, 
academic distribution of instances does not help much to
solve practical cases.

At the beginning of the nineties practitionners in artificial
intelligence realized that classes of random constrained satisfaction problems
used as artificial benchmarks for search algorithms exhibited abrupt
changes of behaviour when some control parameter were finely tuned
\cite{Mi92}. The most celebrated example was random K-Satisfiability, 
where one looks for a solution to a set of random logical constraints
over a set of Boolean variables. It appeared that, for large sets of
variables,  there was a critical
value of the number of constraints per variable below which there
almost surely existed solutions, and above which solutions were
absent. An important feature was that search algorithms performances
drastically worsened in the vicinity of this critical ratio.

This phenomenon, strongly reminiscent of phase transitions in
condensed matter physics, led to a revival of the
interface between statistical physics and computer science, which has
not vanished yet. The purpose of the present lecture is to introduce
the non specialist reader to the concepts and techniques required to
understand the literature in the field. For the sake of simplicity 
the presentation will be limited to one 
computational problem, namely, linear systems of Boolean equations. 
A good reason to do so is that this problem concentrates most of the features 
encountered in other optimization problems, while being 
technically simpler to study. In addition it is closely related to
error-correcting codes in communication theory, see lectures by
A. Montanari and R. Urbanke in the present book.
Extension to other problems will be mentioned in the conclusions.

The lecture is divided into three parts. Sections 1 and 2 are
devoted to the presentation of the model and of elementary concepts 
related to phase transitions e.g. finite-size scaling, large
deviations, critical exponents, symmetry breaking, ... Sections 3
and 4 expose the specific statistical mechanics techniques and concepts
developed in disordered systems to deal with highly interacting and
random systems, namely the replica and cavity approaches. Finally
Section 5 focuses on dynamics and the study of search
algorithms.

\subsection{Linear systems of Boolean equations}

Linear systems of Boolean equations look very much like their
well known counterparts for integer-valued variables, except that
equalities are defined modulo two. 
Consider a set of $N$ Boolean variables $x_i$ with indices $i=1,\ldots
,N$. Any variable shall be False (F) or True (T). The sum of two variables,
denoted by $+$, corresponds to the logical exclusive OR between these
variables defined through,
\begin{eqnarray} \label{sumrule}
F + T &=& T + F = T \quad , \nonumber \\ 
F + F &=& T + T = F  \quad .
\end{eqnarray}
In the following we shall use an alternative representation of the above 
sum rule. Variables will be equal to 0 or 1, instead of $F$ or $T$ 
respectively. Then the $+$ operation 
coincides with the addition between integer numbers modulo two. 

The following is a linear equation involving three variables,
\begin{equation} \label{xoreq01}
x_1 + x_2 + x_3 = 1 \quad .
\end{equation}  
Four among the $2^3=8$ assignments of $(x_1,x_2,x_3)$ satisfy 
the equation: $(1,0,0)$, $(0,1,0)$, $(0,0,1)$ and $(1,1,1)$.
A Boolean system of equations is a set of Boolean equations
that have to be satisfied together. For instance,
the following Boolean system involving four variables 
\begin{equation} \label{xoreq02}
\left\{ \begin{array} {l}
x_1 + x_2 + x_3 = 1 \\
x_2 + x_4  = 0 \\
x_1 + x_4 = 1 
\end{array} \right.
\end{equation}  
has two solutions: $(x_1,x_2,x_3,x_4)=(1,0,0,0)$ and $(0,1,0,1)$. 
A system with one or more solutions is called satisfiable. A trivial
example of an unsatisfiable Boolean system is
\begin{equation} \label{xoreq03}
\left\{ \begin{array} {l}
x_1 + x_2 + x_3 = 1 \\x_1 + x_2 + x_3 = 0 
\end{array} \right. \qquad .
\end{equation}  
Determining whether a Boolean system admits 
an assignment of the Boolean variables satisfying all the
equations constitutes the XORSAT 
(exclusive OR Satisfaction) problem. In the following,
we shall restrict for some reasons to be clarified in Section~\ref{secbasic}
to K-XORSAT, a variant of XORSAT where each Boolean equation include $K$
variables precisely.

K-XORSAT belongs to the class P of polynomial problems \cite{Pa83}. 
Determining whether a system is satisfiable or not 
can be achieved by the standard Gaussian 
elimination algorithm in a time (number of elementary
operations) bounded from above by some constant times 
the cube of the number of bits necessary to store the
system\footnote{The storage space is $K$ times the number of equations
  times the number of bits necessary to label a variable, that is, the
logarithm of the number of variables appearing in the system.}\cite{Pa83}. 

If the decision version of K-XORSAT is easy its optimization version
is not. Assume you are given a system $F$, run the Gauss procedure and
find that it is not satisfiable. Determining the maximal number
$M_S(F)$ of satisfiable equations is a very hard problem. Even 
approximating this number is very hard. It is known that there is no
approximation algorithm (unless P=NP) for XORSAT with ratio $r>\frac
12$, that is, guaranteed to satisfy at least $r\times M_S(F)$ 
equations for any $F$. But $r=\frac 12$ is achieved, on average, by
making a random guess\footnote{Any equation is satisfied by half of
  the configurations of a variables, so a randomly chosen
  configuration satisfies on average $\frac M2 \ge \frac {M_S(F)}2$
equations.}! 

\subsection{Models for random systems}\label{secmodel}

There are many different ways of generating random Boolean systems.
Perhaps the simplest one is the following, called
{\em fixed-size ensemble}. To build an equation we pick up 
uniformly at random $K$ distinct indices among the $N$ ones, say,
${i_1},{i_2}$ and  ${i_k}$. Then we consider the equation
\begin{equation}
x_{i_1} + x_{i_2} + \ldots + x_{i_k} = v \ .
\end{equation}
The second member, $v$, is obtained by tossing a coin: 
$v=0$ or $v=1$ with equal probabilities (one half) and independently of the 
indices of the variables in the first member. 
The process is repeated $M$ times, without correlation between 
equations to obtain a system with $M$ equations.

Another statistical ensemble is the {\em fixed-probability ensemble}.
One scans the set of all $H=2 {N\choose K}$  equations one after the other.
Each equation is added to the system with probability $p$, discarded with
probability $1-p$. Then a system with, on average, $p\,H$ equations
(without repetition) is obtained. In practice one chooses $p=\frac MH$ to
have the same (average) number of equations as in the fixed-size 
ensemble.

The above distributions are not the only possible
ones. However they are easy to implement on a computer, are 
amenable to mathematical studies, and last but not least, 
lead to a surprisingly rich phenomenology. One of the key 
quantities which exhibits an interesting behaviour is
\begin{eqnarray}
P_{SAT} (N,\alpha) &=& \mbox{Probability that a system of random
  K-XORSAT with} \nonumber \\
&& \mbox{ $N$ variables and $M=\alpha\, N$ equations is satisfiable}
 \nonumber \ ,
\end{eqnarray}
which obviously depends on $K$ and the statistical ensemble.
Given $N$ $P_{SAT}$ is a decreasing function of $\alpha$.  
We will see that, in the infinite size limit (and for $K\ge 2$), 
the decrease is abrupt
at some well defined ratio, defining a phase transition between Satisfiable
and Unsatisfiable phase \cite{Cr99}. 
The scope of the lecture is to give some tools
to understand this transition and some related phenomena.


\section{Basic concepts: overview of static phase transitions in K-XORSAT} 
\label{secbasic}

In this Section we introduce the basic concepts necessary to the
study of random K-XORSAT. It turns out that even the $K=1$ case,
trivial from a computer science point of view (each equation
contains a single variable!), can be used as an illustration to
important concepts such as scaling and self-averageness. Ideas 
related to the percolation phase transition and random graphs are
illustrated on the $K=2$ case. Finally the
solution space of 3-XORSAT model exemplifies the notion of clusters
and glassy states.

\subsection{Finite-size scaling (I): scaling function}\label{secsf1}

\begin{figure}[t]
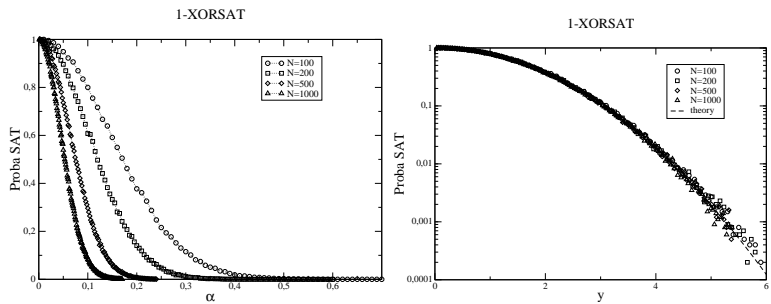

\includegraphics*[width=5.truecm,angle=0]{fig211.eps}
\includegraphics*[width=5.truecm,angle=0]{fig212.eps}
\caption{Left: Probability that a random 1-XORSAT formula is satisfiable as
  a function of the ratio $\alpha$ of equations per variable, and for
  various sizes $N$. Right: same data as in the left panel after the
horizontal rescaling $\alpha \to \alpha \times \sqrt N = y$; note the use of
  a log scale for the vertical axis. The dashed line
  shows the scaling function $\Phi_1(y)$ (\ref{sca1}).}
\label{proba1-fig}
\end{figure}

Figure \ref{proba1-fig}(left) shows the probability $P_{SAT}$ that a randomly
extracted 1-XORSAT formula is satisfiable as a function of the ratio
$\alpha$, and for sizes $N$ ranging from 100 to 1000. We see that
$P_{SAT}$ is a decreasing function of $\alpha$ and $N$. 

Consider the subformula made of the $n_i$ equations with first member equal to 
$x_i$. This formula is always satisfiable if $n_i=0$ or $n_i=1$. If
$n_i\ge 2$ the formula is
satisfiable if and only if all second members are equal (to 0, or to
1), an event with probability $(\frac 12)^{n_i-1}$ 
decreasing exponentially with the number
of equations. Hence we have to consider the following variant of the celebrated
Birthday problem\footnote{The Birthday problem is a classical
elementary  probability problem: given a class with $M$ students, what is the
  probability that at least two of them have the same birthday date?
  The answer for $M=25$ is $p\simeq 57\%$, while a much lower value is
  expected on intuitive grounds when $M$ is much smaller than the number
$N=365$   of days in a year.}. Consider a year with a number $N$
of days, how should scale the number $M$ of students in a 
class to be sure that no two students have the same birthday date?
\begin{equation}
\bar p =\prod_{i=0}^{M-1}\left(1-\frac iN\right) = \exp
\left(-\frac{M(M-1)}{2N} + O(M^3/N^2) \right) \ .
\end{equation}
Hence we expect a cross-over from large to small $\bar p$ when $M$
crosses the scaling regime $\sqrt N$. Going back to the 1-XORSAT model
we expect $P_{SAT}$ to have a non zero limit value when the number of
equations and variables are both sent to infinity at a fixed ratio
$y=M/\sqrt N$. In other words, random 1-XORSAT formulas with $N$
variables, $M$ equations or with, say, $100\times N$ variables,
$10\times M$ equations should have roughly the same probabilities of 
being satisifiable. To check this hypothesis we replot the data in 
Figure~\ref{proba1-fig} after multiplication of the abscissa of each
point by $\sqrt N$ (to keep $y$ fixed instead of $\alpha$). The
outcome is shown in the right panel of Figure~\ref{proba1-fig}. Data
obtained for various sizes nicely collapse on a single limit curve
function of $y$.

The calculation of this limit function, usually called scaling
function, is done hereafter in the fixed-probability
1-XORSAT model where the number of equations
is a Poisson variable of mean value $\bar M=y\sqrt N$.  We will
discuss  the equivalence between the fixed-probability
and the fixed-size ensembles later.
In the fixed-probability ensemble the numbers $n_i$ of occurence of each
variable $x_i$ are independent Poisson variables with average value
$\bar M/N=y/\sqrt N$. Therefore the probability of satisfaction is 
\begin{eqnarray} \label{psatp}
P_{SAT} ^{p} (N,\alpha=\frac y{\sqrt N}) 
&=& \left[  e^{- y/\sqrt N} \left( 1+ \sum _{n\ge 1}
\frac {(y/\sqrt N)^n}{n!} \left(\frac 12\right)^{n-1}\right) \right] ^N  
\nonumber \\ &=& \left[ 2 e^{-y/(2\sqrt N)}-e^{-y/\sqrt N}\right]^N  \ , 
\end{eqnarray}
where the $p$ subscript denotes the use of the fixed-probability ensemble.
We obtain the desired  scaling function
\begin{equation} \label{sca1}
\Phi _1(y)\equiv \lim _{N\to\infty} \ln P^p_{SAT} (N,\alpha=\frac y{\sqrt
  N}) = -\frac{y^2}4 \ ,
\end{equation}
in excellent agreement with the rescaled data of
Figure~\ref{proba1-fig} (right) \cite{Cr03a}. 
    
\subsection{Self-averageness of energy and entropy}\label{secselfav}

Let us now consider random 1-XORSAT formulas at a finite ratio
$\alpha$, and ask for the distribution of the minimal fraction of
unsatisfied equations, hereafter called ground state (GS) energy $e_{GS}$. 
For simplicity we work in the fixed-probability ensemble again. The
numbers $n_i^0, n_i^1$  of, respectively, $x_i=0, x_i=1$ are
independent Poisson variables with mean $\frac \alpha 2$. The
minimal number of unsatisfied equations is clearly $\min (n_i^0,
n_i^1)$. The GS energy is the sum (divided by $M$) of $N$ such 
i.i.d. variables; from the law of large number it almost surely 
converges towards the average value
\begin{equation}
e_{GS} (\alpha) = \frac 12 \left( 1 - e^{-\alpha} I_0(\alpha) -
e^{-\alpha} I_1(\alpha) \right) \ ,
\end{equation}
where $I_\ell$ denotes the $\ell^{th}$ modified Bessel function.
In other words almost all formulas have the same GS energy in the
infinite $N$ limit, a property called self-averageness in physics, and
concentration in probability. 

How many configurations of variables realize have minimal energy?
Obviously a variable is free (to take 0 or 1 value) if $n_i^0=n_i^1$,
and is frozen otherwise. Hence the number of GS configurations is
${\cal N} = 2^{N_f}$ where $N_f$ is the number of free variables. 
Call 
\begin{equation}
\rho = \sum _{n\ge 0} e^{-\alpha} \left(
\frac{\alpha} 2\right)^n \frac 1{(n!)^2} = e^{-\alpha}\; I_0(\alpha)
\end{equation}
the probability that a variable is free. Then $N_f$ is a binomial
variable with parameter $\rho$ among $N$; it is sharply concentrated
around $\overline{N_f}=\rho\, N$ with typical fluctuations of the
order of $N^{1/2}$. As a consequence, the GS entropy per variable, 
$s_{GS} = (\log {\cal N})/N$, is self-averaging and almost surely
equal to its average value $s_{GS} = \rho \log 2$.

Self-averageness is the very useful property. It allows us to study
the average value of a random variable, instead of its full
distribution. We shall use it in Section \ref{secreplicas} 
and also in the analysis of algorithms of Section \ref{secuc}. 
This property is not restricted to XORSAT but was proven to 
hold for the GS energy \cite{Br93} and entropy \cite{Mo07} of other
optimization problems.  

Not all variables are self-averaging of course. A straightforward
example is the number ${\cal N}$ of GS configurations itself. 
Its $q^{th}$ moment reads
$\overline{{\cal N}^q} = \left(1-\rho + \rho \; 2^q\right)^N$ 
where the overbar denotes the average over the formulas. We see that
$\overline{{\cal N}^q} \gg (\overline{{\cal N}})^q$: ${\cal N}$
exhibits large fluctuations and is not concentrated around its
average. Very rare formulas with atypically large number
$N_f$ of free variables contribute more to the $q^{th}$ moment than
the vast majority of formulas, and spoil the output. This is the very
reason we will need the introduction of the replica approach in
Section \ref{secreplicas}.

\subsection{Large deviations for $P_{SAT}$ (I): 1-XORSAT}\label{seclargedev}

As we have seen in the previous sections 1-XORSAT formulas with a
finite ratio $\alpha$ are unsatifiable with high probability {\em
  i.e.} equal to unity in the infinite $N$ limit. For finite but large
$N$ there is a tiny probability that a randomly extracted formula is 
actually satisifiable. A natural question is to characterize the
`rate' at which $P_{SAT}$ tends to zero as $N$ increases (at fixed
$\alpha$). Answering to such questions is the very scope of large
deviation theory (see \ref{applargedev} for an elementary
introduction). Looking for events with very small probabilities 
is not only interesting from an academic point of view, but can also
be crucial in practical applications. We will see in Section 
\ref{secescape} that
the behaviour of some algorithms is indeed dominated by rare events.

\begin{figure}[t]
\includegraphics*[width=5.truecm,angle=0]{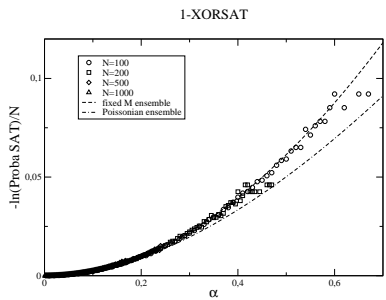}
\caption{Same data as Figure\ref{proba1-fig} (left) with: logarithmic
  scale on the vertical axis, and rescaling by $-1/N$. The scaling
  functions $\omega_1$ (\ref{sca2})
and $\omega_1 ^p$ (\ref{sca2p}) for, respectively, the fixed-size
  and fixed-probability ensembles are shown.}
\label{proba1b-fig}
\end{figure}

Figure~\ref{proba1b-fig} shows minus the logarithm of $P_{SAT}$, divided by
$N$, as a function of the ratio $\alpha$ and for various sizes
$N$. Once again the data corresponding to different sizes collapse on
a single curve, meaning that
\begin{equation}
P_{SAT} (N,\alpha) = e^{-N\; \omega _1(\alpha) + o(N)} \ .
\end{equation}
Decay exponent $\omega_1$ is called rate function in probability
theory. We can derive its value in the fixed-probability ensemble from 
(\ref{psatp}) with $y=\alpha\times \sqrt N$, with the immediate result 
\begin{equation} \label{sca2p}
\omega_1 ^p (\alpha) = \alpha - \ln \big( 2\; e^{\alpha/2} -1 \big) \ .
\end{equation}
The agreement with numerics is very good for small ratios,
but deteriorates as $\alpha$ increases. The reason is simple. In the
fixed-probability ensemble the number $M$ of equations is not fixed but may
fluctuate around the average value $\bar M =\alpha N$. The 
ratio $\tilde \alpha =M/N$, is with high probability equal to
$\alpha$, but large deviations ($\tilde \alpha \ne \alpha)$ are
possible and described  by the rate
function\footnote{$M$ obeys a Poisson law with parameter $\bar
  M$. Using Stirling formula, $$e^{-\bar M} \frac{\bar M ^M}{M!}
  \simeq e^{-\alpha N } (\tilde \alpha N)^{\alpha N}{\sqrt {2\pi N}} 
\left( \frac e{\alpha N}\right)^{\alpha N} = e^{-N \, \Omega
  (\tilde\alpha | \alpha) + o(N)}\ ,$$ where $\Omega$ is defined in
(\ref{defOmega}). },
\begin{equation}\label{defOmega}
\Omega (\tilde \alpha | \alpha ) = \tilde \alpha -\alpha - \alpha\;
\ln (\alpha/\tilde \alpha ) \ .
\end{equation}
However the probability that a random 1-XORSAT formula with $M$
equations is satisfiable is also exponentially small in $N$, with a
rate function $\omega_1(\alpha)$ increasing with $\alpha$. Thus, in
the fixed-probability ensemble, a trade-off is found between ratios $\tilde
\alpha$ close to $\alpha$ (formulas likely to be generated) and close
to 0 (formulas likely to be satisfiable). As a result 
the fixed-probability rate function is 
\begin{equation} \label{sca2}
\omega_1 ^p (\alpha) = \min _{\tilde \alpha } \big[ \omega _1 (\tilde
  \alpha) + \Omega (\tilde \alpha | \alpha )\big] \ ,
\end{equation}
and is smaller than $\omega_1(\alpha)$. 
It is an easy check that the optimal ratio
$\tilde \alpha ^*=\alpha /(2-e^{-\alpha/2}) < \alpha$ as
expected. Inverting (\ref{sca2}) we deduce the rate function
$\omega_1$ in the
fixed-size ensemble, in excellent agreement with numerics
(Figure~\ref{proba1b-fig}). This example underlines that 
thermodynamically equivalent ensembles
have to be considered with care as far as rare events are concerned.

Remark that, when $\alpha \to 0$, $\tilde \alpha =\alpha +
O(\alpha ^2)$, and $\omega _1 ^p(\alpha)=\omega_1 (\alpha)
+O(\alpha^3)$.  This common value coincides with the scaling
function $- \Phi_1(\alpha)$ (\ref{sca1}). This identity is expected on
general basis (Section \ref{secld2}) and justifies the agreement between the
fixed-probability scaling function and the numerics based on the fixed-size
ensemble (Figure~\ref{proba1-fig}, right).

\subsection{Percolation in random graphs}\label{secrg}

Though 1-XORSAT allowed us to understand some general features of random
optimization problems it is very limited due to the absence of interactions 
between variables. A more interesting problem is 2-XORSAT where 
every equation define a joint constraint on two variables. 
Formulas of 2-XORSAT can be represented by a graph with $N$ vertices
(one for each variable), and $\alpha N$ edges. 
To each equation of the type $x_i+x_j=e$
corresponds an edge linking vertices $i$ and $j$, and carrying
0 or 1 label (the value $e$ of the second member).
Depending on the input model chosen (Section \ref{secmodel}) multiple edges 
are present or not. 

\begin{figure}
\begin{center}
\includegraphics[width=70pt,angle=0]{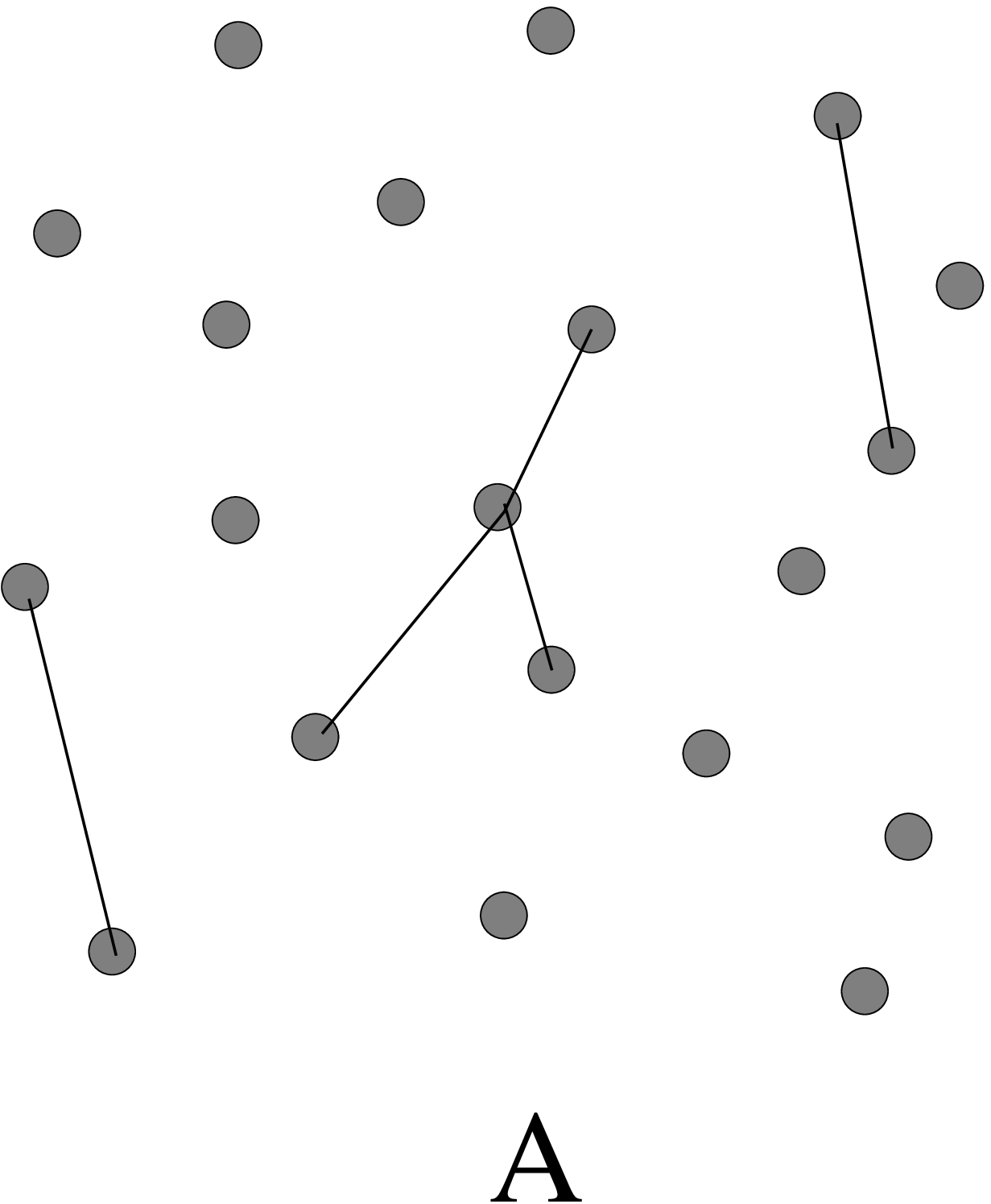} 
\hskip 1cm
\includegraphics[width=70pt,angle=0]{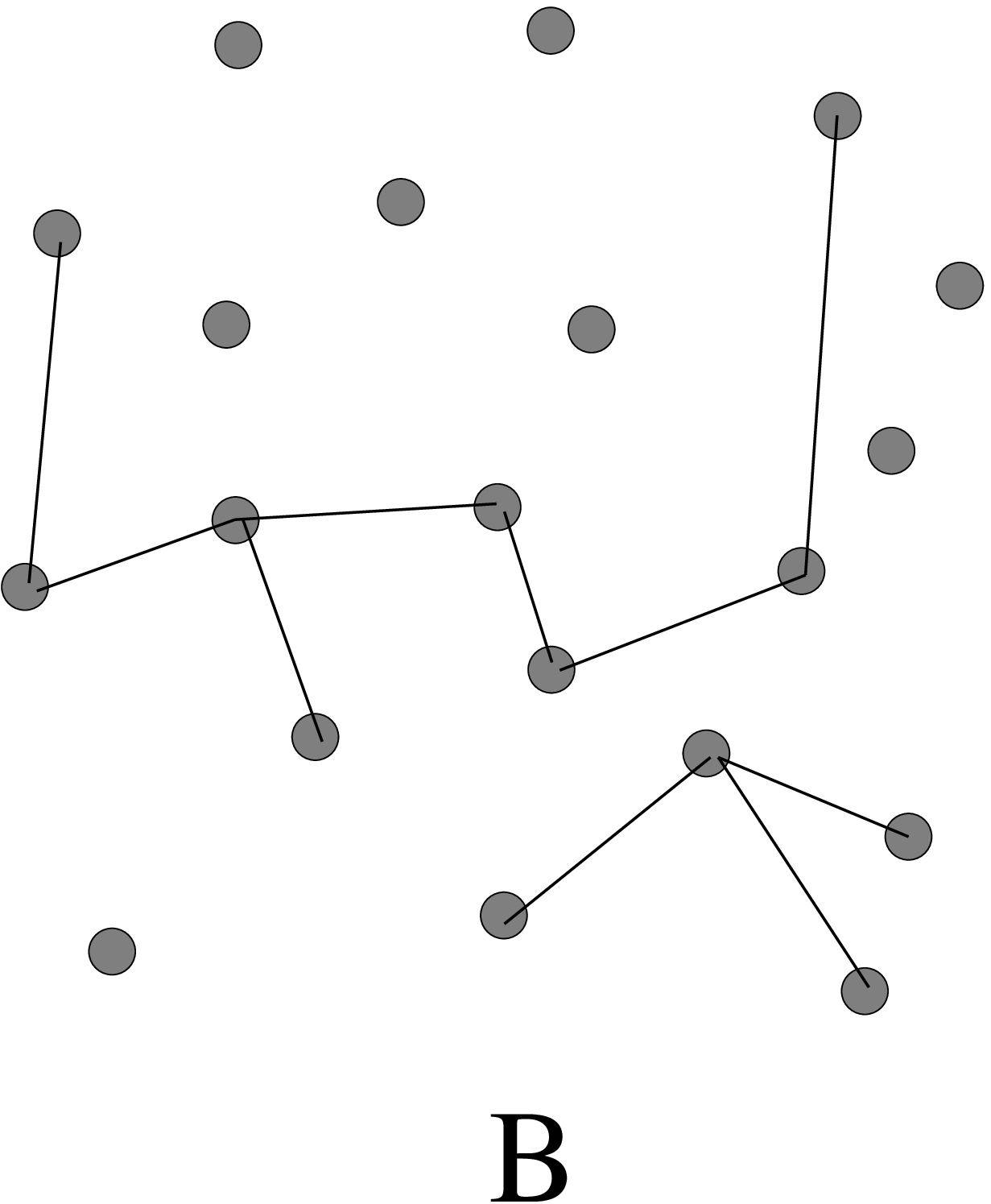}
\hskip 1cm
\includegraphics[width=70pt,angle=0]{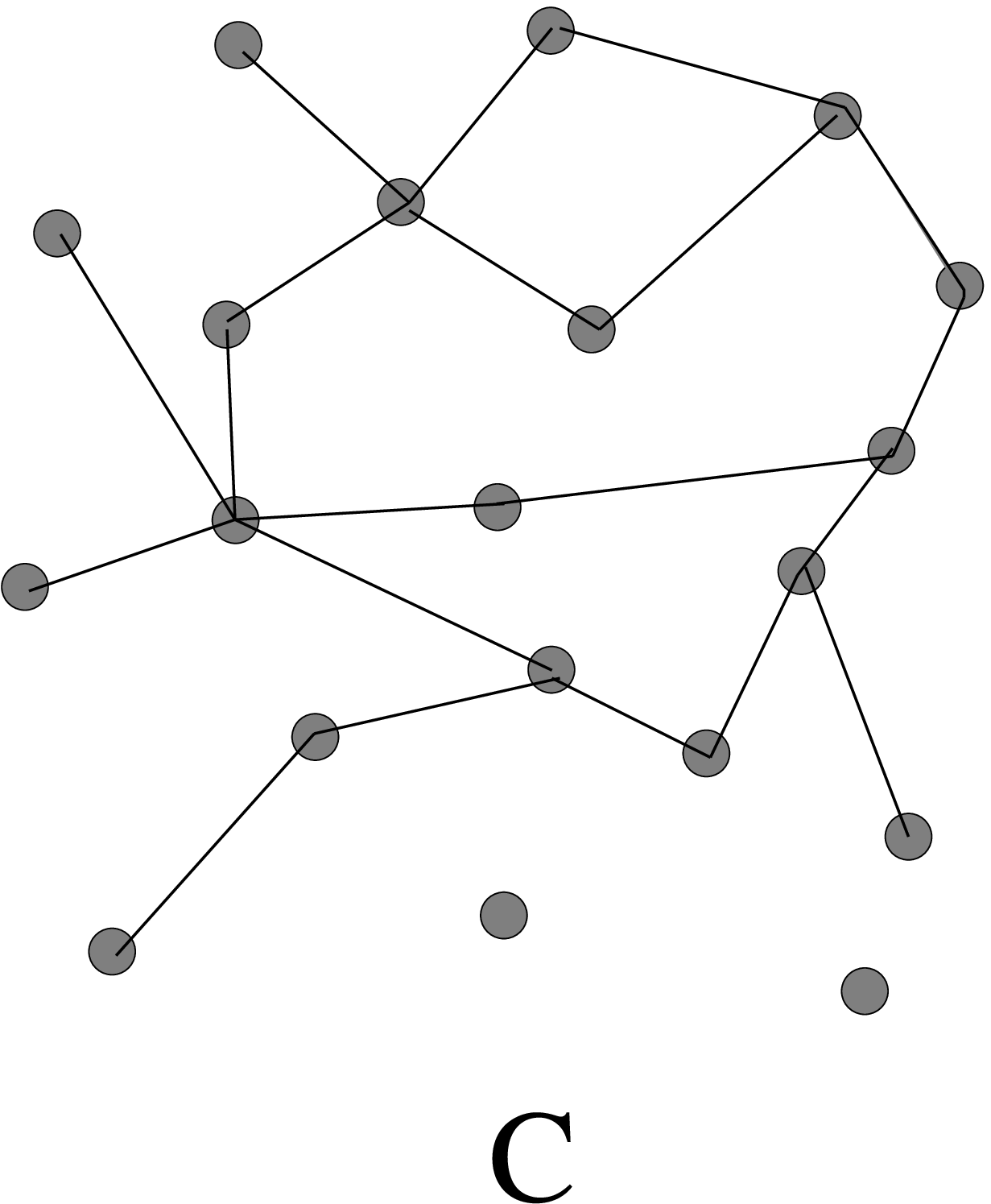}
\caption{Examples of random graphs generated at fixed number $M=\alpha N$ of
edges (fixed-size model without repetition). All graph include $N=20$ 
vertices (grey dots). The average degree of valency, $2\alpha$,
is equal to $0.5$ ({\bf A}),  $1$ ({\bf B}), and
$2$ ({\bf C}). The labels of the vertices have been permuted to
obtain planar graphs, {\em i.e.} avoid crossing of edges.}
\label{graphealeatoire}
\end{center}
\end{figure}

As the formula is random so is graph. 
Figure~\ref{graphealeatoire} shows examples of graphs obtained 
 for various values of $\alpha$. Notice the
qualitative change of structure of graphs when 
the ratio $\alpha$ varies from low values (graphs are mostly made
of small isolated trees) to higher ones (a large part of vertices
are now connected together). This change is known as the percolation
transition in physics, or the appearance of a giant component in
mathematics literature.

Before reviewing some of the aspects of the percolation transition
let us mention an important fact on the valency of vertices.
As a result of the randomness of the graph generation process,
each node share edges with a variable number of neighboring vertices. 
In the large $N$ limit the degree $v$ of a vertex, {\em i.e.} 
the number of its neighbors, 
is a Poisson variable with mean $2\alpha$,
\begin{equation} \label{poissonconnec}
\mbox{Proba}[v] = e^{-2\alpha}\; \frac{(2\alpha)^v}{v!} \ .
\end{equation}
For instance the fraction of isolated vertices is $e^{-2\alpha}$.
The average degree of a vertex, $c=2\alpha$, is called
connectivity. 

It is natural to decompose the graphs into its connected subgraphs,
called components. Erd\"os and R\'enyi were able in 1960 to characterize the 
distribution of sizes of the largest component \cite{Bo89},
\begin{itemize}
\item When $c<1$, the largest component includes
$\sim \ln N/(c-1-\ln c)$  vertices  with high probability. 
Most components include only
a finite number of vertices, and are trees {\em i.e.} contain no circuit.
\item For $c=1$ the largest component contain $O(N^{2/3})$ vertices.
\item When $c>1$ there is one giant component containing $\sim \gamma
(c) N$ vertices; the others components are small {\em i.e.} look
like the components in the $c<1$ regime.
The fraction of vertices in the giant component is the unique
positive solution of 
\begin{equation} \label{sizegiant}
1-\gamma = e^{-c \, \gamma} \ .
\end{equation}
It is a non analytic function of $c$, equal to 0 for $c\le 1$, and 
positive above, tending to unity when $c$ increases.
\end{itemize}

The phenomenon taking place at $c=1$ is an example of (mean-field) 
percolation transition. We now give a hand-waving derivation of
(\ref{sizegiant}).   
Consider a random graph $G$ over $N$ vertices, with connectivity
$c$. Add a new vertex $A$ to the graph to obtain $G'$. If we want $G'$
to be drawn from the same distribution as $G$, a number $v$ of edges
must be attached to $A$, where $v$ an integer--valued random number
following the Poisson distribution (\ref{poissonconnec}). 
After addition of $A$, some connected components of $G$ will merge
in $G'$. In particular, with some probability $p_v$, $A$ will be part
of the giant component of $G'$. To estimate $p_v$, we note that this
event will not happen if and only if none of the $v$ neighbors of $A$ in
$G'$  belongs to the giant component of $G$. Thus,
\begin{equation} \label{processusgraphe}
1-p_v = (1-\gamma ) ^v \qquad ,
\end{equation}
where $\gamma$ is the size (fraction of vertices) of the giant component.
Summing both sides of (\ref{processusgraphe})
over the distribution (\ref{poissonconnec}) for $v$, and
asserting that the change in size of the giant component between
$G$ and $G'$ is $o(1)$ for large $N$, we obtain  (\ref{sizegiant}).

The above derivation illustrates an ubiquitous idea in 
probability and
statistical physics, which could be phrased as follows: `if a system
is very large, its statistical properties should be, in some sense, 
unaffected by a small increase in size'. This idea will be useful, in
a more sophisticated context, in Section \ref{seccavity}.

\subsection{Sat/Unsat transition in 2-XORSAT} \label{secsatunsat2}

\begin{figure}[t]
\includegraphics*[width=5.truecm,angle=0]{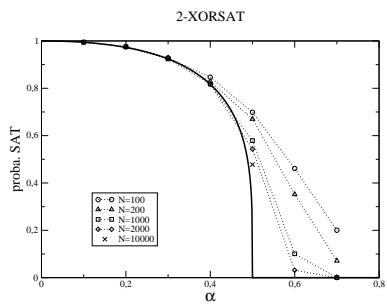}
\caption{Probability that a random 2-XORSAT formula is satisfiable as
  a function of the ratio $\alpha$ of equations per variable, and for
  various sizes $N$. The full line is the asymptotic analytical
formula (\ref{psaty2}).}
\label{proba2-fig}
\end{figure}

Figure \ref{proba2-fig} shows the probability $P_{SAT}$ 
that a randomly extracted 
2-XORSAT formula is satisfiable as function of $\alpha$, and for
various sizes $N$. It appears that $P_{SAT}$  drops quickly 
to zero for large $N$ when $\alpha$ reaches 
the percolation threshold $\alpha_c = \frac 12$. For ratios smaller than
$\alpha _c$ the probability of satisfaction is positive, but smaller 
than unity. 

Take $\alpha < \frac 12$. Then the random graph $G$ associated to a random
2-XORSAT formula is non percolating, and made of many small components. 
Identical components (differing only by a relabelling of
the variables) may appear several times, depending on their topology.
For instance consider a connected graph $G'$ made of $E$ edges and $V$ 
vertices.
The average number of times $G'$ appears in $G$ is a function of $E$ and 
$V$ only, 
\begin{equation}
N_{E,V} = {N \choose V} \left( \frac{2\alpha}N \right)^E 
 \left( 1 - \frac{2\alpha}N \right)^{\frac {V(V-1)}2 + V(N-V)} 
\end{equation}
since any vertex in $G'$ can establish edges with other vertices in
$G'$, but is not allowed to be connected to any of the $N-V$ outside vertices.
When $N$ is very large compared to $E,V$ we have
\begin{equation} \label{avtree}
N_{E,V} \simeq N^{V-E} \; \frac{ (2\alpha)^E}{V!}\;  
e^{-2\alpha\, V} \ .
\end{equation}
Three cases should distinguished, depending on the value of $V-E$:
\begin{itemize}
\item{$V-E=1$}: this is the largest value compatible with connectedness, 
and corresponds to the case of trees. From (\ref{avtree}) every 
finite tree has of the order of $N$ copies in $G$.
\item {$V-E=0$}: this correspond to trees with one additional edge, that is, to graphs
having one cycle (closed loop). The average number of unicyclic graphs is,
from (\ref{avtree}), finite when $N\to\infty$. 
\item{$V-E\le -1$}: the average number of components with more than one cycle 
vanishes in the large $N$ limit; those graphs are unlikely to be
found and can be ignored\footnote{The probability that such a graph exists is
bounded from above by the average number, see \ref{appmoment}.}. 
\end{itemize}
Obviously a 2-XORSAT formula with tree
structure is always satisfiable\footnote{Start from one leaf, assign the 
attached variable to 0, propagate to the next variable according to
the edge value, and so on, up to the completion of the tree.}. Hence
dangerous subformulas, as far as satisfiability is concerned, are
associated to unicyclic graphs. A simple thought shows that a unicyclic
formula is
satisfiable if and only if the number of edges carrying label 1
along the cycle is
even. Since the values attached to the edges (second members in the 
formula) are uncorrelated 
with the topology of the subgraph (first members)
each cycle is satisfiable with probability one half.
We end up with the simple formula
\begin{equation} \label{psatx2}
P_{SAT} (N,\alpha) = \langle 2 ^{- C (G)} \rangle 
\end{equation}
where $C(G)$ denotes the number of cycles in $G$, and 
$\langle .\rangle$ the average over $G$.  For a reason which will become
clear below let us classify cycles according to their length $L$. 
How many cycles of length $L$ can we construct? We have to choose first
$L$ vertices among $N$, and join them one after the order according
to some order. As 
neither the starting vertex nor the direction along the cycle matter, 
the average number of $L$-cycles is
\begin{equation} \label{deflaml}
N_L =\frac{ N(N-1)\ldots (N-L+1)}{2L} \times \left( \frac{2\alpha}N \right)^L 
\to \Lambda _L = \frac{(2\alpha)^L}{2L}\ .
\end{equation}
when $N\to\infty$. As the emergence of a cycle between $L$ vertices is a 
local event (independent of the environment) we expect the number of
$L$-cycles to be Poisson distributed in the large $N$ limit
with parameter $\Lambda_L$. This statement can actually be proven, and
extended to any finite collection of cycles of various lengths\cite{Bo89}:
in the infinite size limit, the joint distribution of the numbers of cycles
of lengths $1,2, \ldots, L$ is the product of Poisson laws with
parameters $\Lambda_1, \Lambda _2, \ldots,\Lambda _L$
calculated in (\ref{deflaml}). The probability of satisfaction 
(\ref{psatx2}) therefore converges to 
\begin{equation}
\lim _{N\to\infty} P_{SAT} (N,\alpha) 
= \prod_{L\ge L_0}  \left\{ \sum _{C\ge 0} 
e^{-\Lambda _L} \; \frac{{(\Lambda _L/2)}^C}{C!} 
\right\} = \prod _{L\ge L_0} e^{-\Lambda_L/2}
\end{equation}
where $L_0$ is the minimal cycle length. In normal random graphs $L_0=3$ 
since triangles are the shortest cycles. However in our 2-XORSAT model
any equation, or more precisely, any first member  can appear twice or 
more, hence $L_0=2$. We conclude that \cite{Cr03a}
\begin{equation} \label{psaty2}
\lim _{N\to\infty} P_{SAT} (N,\alpha) 
= e^{\alpha/2}\; (1-2\alpha) ^{\frac 14} \qquad 
\mbox{when} \qquad \alpha < \alpha_c =\frac 12 \ .
\end{equation}
The agreement of this result  
with the large size trend coming out from numerical 
simulations is visible in Figure \ref{proba2-fig}. As $P_{SAT}$ is a 
decreasing function of $\alpha$ it remains null for all ratios larger than 
$\alpha_c$. The non analyticity of $P_{SAT}$ at $\alpha_c$
locates the Sat/Unsat phase transition of 2-XORSAT. 

It is an implicit assumption of statistical physics that asymptotic 
results of the kind of (\ref{psaty2}), rigorously valid in the $N\to
\infty$ limit, should reflect with good accuracy the finite but large
$N$ situation. An inspection of Figure \ref{proba2-fig} shows this is 
indeed the case. For instance, for ratio $\alpha =.3$, (\ref{psaty2}) 
cannot be told from the probability of satisfaction 
measured for formulas with $N=100$ variables. This statement
does not hold for $\alpha =.4$, where the agreement between infinite size
 theory and numerics sets in when $N = 1000$ at least. It appears
that such finite-size effects become bigger and bigger as $\alpha$
gets closer and closer to the Sat/Unsat threshold. This issue, of broad 
importance in the
context of phase transitions and the pratical application of asymptotic
results, is studied in Section \ref{secfs2}.

\subsection{Large deviations for $P_{SAT}$ (II): bounds in
the Unsat phase of 2-XORSAT.} \label{secld2}

Consider ratios $\alpha > \alpha_c$. 
The giant components of the corresponding formulas  contain an 
extensively large number of independent cycles, so we expect 
from  (\ref{psatx2}) that the probability of satisfaction is 
exponentially small in $N$, $P_{SAT} =\exp (-N \omega_2(\alpha)+o(N))$. 
Lower and upper bounds to the rate function $\omega _2$ can be obtained 
from, respectively, the first and second moment inequalities 
described in \ref{appmoment}. Denoting by ${\cal N}$ the number of solutions
of a formula $P_{SAT}$ is the probability that ${\cal N}\ge 1$, and
is bracketed according to (\ref{ineq}). 

To calculate the first moment of ${\cal N}$ remark that an equation
is satisfied by one half of the configurations. This result remains true 
for a restricted set of configurations when we average over the possible
choices of (the second member of) the equation. The average number
of solutions is thus $2^N/2^M$, from which we get
\begin{equation} \label{lowerom2}
\omega _2 (\alpha) \ge (\alpha -1) \ln 2 \ .
\end{equation}
This lower bound is useless for $\alpha <1$ since $\omega_2$ is positive
by definition. As for the upper bound we need to calculate the 
second moment $\langle {\cal N}^2\rangle$ of ${\cal N}$. As equations are 
independently drawn 
\begin{equation} \label{n2}
\langle {\cal N} ^2\rangle = \sum _{X,Y} q (X,Y) ^M \, 
\end{equation} 
where the sum is carried out over the pairs $X,Y$ of configurations
of the $N$ variables, and $q(X,Y)$ is the
probability that both $X$ and $Y$ satisfies the same randomly
drawn equation. $q$ can be easily expressed in terms of 
the Hamming distance $d$ between
$X$ and $Y$, defined as the fraction of variables having opposite
values in $X$ and $Y$. The general
expression for K-XORSAT is\footnote{The
equation is satisfied if the number of its variables taking opposite
values in $Y$ as in $X$ is even.
definition of $d$ the probability (over its index $i$) that a variable
takes different value in $X$ and $Y$ is $d$. Hence expression
(\ref{qk})
for $q(d)$. Beware of  the  $O(\frac 1N)$ corrections to this expression
e.g. if variable $x_1\ne y_1$ (which happens with probability $d$) then
the probability that   $x_2\ne y_2$ is $(dN-1)/(N-1) = d + (1-d)/-N-1)$. 
Those corrections are relevant for the calculation of Gaussian
fluctuations around the saddle-point (\ref{appfluctu}).}  
\begin{equation} \label{qk}
q  (d)=\frac 12( 1 - (1-2d)^K) \,
\end{equation}
and we specialize in this section to $K=2$.
Going back to (\ref{n2}) we can sum over $Y$ at fixed $X$, that is,
over the distances $d$ taking multiple values of $\frac 1N$ with the 
appropriate binomial multiplicity, and then sum over $X$ with the result
\begin{equation} \label{n2p}
\langle {\cal N} ^2\rangle = 2^{N} \sum _{d} {N \choose N\,d} \; q(d)^M
= \exp( N\, \max_{d \in [0;1]}A(d,\alpha) +o(N))
\end{equation} 
in the large $N$ limit, where
\begin{equation} \label{upperom2}
A(d,\alpha) =  (2\alpha-1) \; \ln 2 
  -d \ln d -(1-d) \ln (1-d) +\alpha \ln  q(d)  \ .
\end{equation}
For $\alpha < \frac 12$ the maximum of $A$ is located in $d^*=\frac 12$, 
and equal to $A^*=0$.  When $\alpha > \frac 12$,  $A$ has two global 
maxima located in $d^*(\alpha) <\frac 12$ and $1-d^*(\alpha)$, with 
equal value $A^*(\alpha) >0$.  

\begin{figure}[t]
\includegraphics*[width=5.truecm,angle=0]{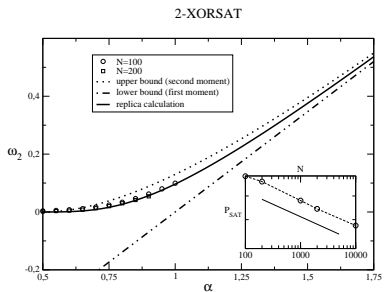}
\caption{Rate function $\omega_2 (\alpha)$ associated to the probability
of satisfaction of 2-XORSAT formulas with ratio $\alpha$. The dotted line
is the upper bound (\ref{upperom2}) and the dot-dashed line the lower bound
(\ref{lowerom2}). The full line is the output of the replica calculation of
Section \ref{secresu}, squares and circles represent numerical results for
$N=200$, 100 respectively from $10^6$ formulas. 
Inset: $P_{SAT}$ as a function of the size $N$ at the
Sat/Unsat ratio. The slope $-\frac 1{12}$ (\ref{predic112}) is shown for 
comparison.} 
\label{proba2b-fig}
\end{figure}

We plot in  Figure \ref{proba2b-fig} the lower (\ref{lowerom2}) and
upper bounds to the rate function, 
\begin{equation} \label{om2}
\omega _2 (\alpha) \le 2(1-\alpha)\ln 2 - \max_{d \in [0;1]}A(d,\alpha)
\end{equation}
from  (\ref{ineq}). 
At large ratio both bounds asymptotically match, proving that
$\omega_2 (\alpha) = (\alpha-1) \ln 2 + O(e^{-2\alpha})$. 
As the ratio departs from its threshold value by 
$\epsilon = \alpha - \alpha _c$ the upper bound grows quadratically,  
 $A^*(\alpha_c + \epsilon )\simeq \frac 34 
\epsilon^2 + O(\epsilon ^3)$. Numerics suggest that  the increase of
the rate function is slower, 
\begin{equation} \label{om2clth}
\omega _2 (\alpha_c + \epsilon) \simeq \Omega\;   \epsilon ^3 +
O(\epsilon ^4)\ ,
\end{equation} 
for some constant $\Omega\simeq 1$ (Figure \ref{proba2b-fig}). 
We will see in Section \ref{secreplicas} that
a sophisticated statistical physics technique, called the replica
method, actually predict this scaling with $\Omega = \frac {32}{27}$.
Actually the rate function can be estimated with the replica approach
for any ratio $\alpha$ with the result shown in Figure \ref{proba2b-fig}.

\subsection{Order parameter and symmetry breaking}\label{secop}

What is the meaning of the Hamming distance $d^*(\alpha)$ appearing
in the calculation of the second moment of the number of solutions?
An easy guess would be the average distance between pairs
of solutions
\begin{equation} \label{dav}
 d_{av}(\alpha) = \lim _{N\to\infty} \langle \frac{\sum _{X,Y\ \mbox{solutions
of}\ F} d(X,Y)}{{\cal N}(F)^2}\rangle_F
\end{equation}
where the average is taken over the satisfiable 
formulas $F$ with ratio $\alpha$, and
$d(X,Y)$ denotes the (intensive) Hamming distance between two 
solutions $X,Y$. However an inspection of the calculation of Section 
\ref{secld2} shows that
\begin{equation} \label{davstar}
 d^*(\alpha) = \lim _{N\to\infty} \frac{\langle \sum _{X,Y\ \mbox{solutions
of}\ F} d(X,Y)\rangle _F}{\langle {\cal N}(F)^2 \rangle_F} \ne 
d_{av}(\alpha)\ .
\end{equation}
Actually, though $d^*(\alpha)$ is not the average distance
between solutions with the unbiased distribution over  formulas, it 
is the average distance for a biased distribution  where 
each formula is weighted with
\begin{equation}\label{biased}
w(F) = \frac{  {\cal N}(F)^2}{\sum_{F'}  {\cal N}(F')^2}
\end{equation}
as can be readily checked upon insertion of $w(F)$ in the numerator
of (\ref{dav}). We will see in Section (\ref{secreplicas}) how 
to calculate average properties with the unbiased measure. 

Even so definition (\ref{davstar}) (and (\ref{dav}) too) is sloppy. 
If $X$ is a solution so is $-X$, the configuration where variables
values are flipped. Thus the average distance, whatever the weights
over formulas, is equal $\frac 12$ for any $N$! The difficulty comes from
the ambiguity in how the thermodynamic limit is taken, and
is the signature of spontaneous symmetry breaking. In the
low temperature phase of the Ising model 
the magnetization is either $m^*>0$ or $-m^*<0$ 
if an external field $h$ with, respectively, positive or negative vanishing  
amplitude is added prior to taking the infinite size limit.
In the present case what plays the role of the field  is a coupling
between solutions as is well-known in splin-glass theory 
\cite{Pa86}. Inserting $\exp[- N \, h\, d(X,Y)]$ in the numerator
of (\ref{davstar}) we obtain, when $N\to\infty$, $d^*$ if $h \to 0^+$
and $1-d^*$ if $h\to 0^-$.  The density $\mu$ of probability
of distances $d$ between solutions, with the biased measure (\ref{biased}), 
is concentrated below the Sat/Unsat threshold,
\begin{equation}
\mu (d) = \delta\big( d - \frac 12 \big) \qquad \mbox{for} \qquad
\alpha <\alpha _c \ , 
\end{equation}
and split into two symmetric peaks above the critical ratio,
\begin{equation}
\mu (d) = \frac 12 \delta\big( d - d^* \big) +
\frac 12 \delta\big( d - (1-d^*) \big)\qquad \mbox{for} \qquad
\alpha >\alpha _c \ . 
\end{equation}
The concept of spontaneous symmetry breaking will play a key role
in our study of 3-XORSAT (Section \ref{secpin}).

\subsection{Finite-size scaling (II): critical exponents}
\label{secfs2}

Let us summarize what we have found about the probability of satisfying 
random 2-XORSAT formulas in Section \ref{secsatunsat2} 
and \ref{secld2}. Close to the transition we have from (\ref{psaty2}) and 
(\ref{om2clth}),
\begin{equation} \label{bornesp2}
\ln P_{SAT} (N,\alpha_c +\epsilon) \simeq \left\{ \begin{array} {c c c}
\frac 14 \ln (-\epsilon) & \mbox{when} & \epsilon<0, N\to \infty\\
- \Omega\; N \; \epsilon ^3  & \mbox{when} & \epsilon>0, N\gg 1 \nonumber
\end{array}\right . \quad .
\end{equation}
The lesson of Section \ref{secsf1} is that $\ln P_{SAT}$ may 
have a non trivial limit  when $N\to \infty$,  $\epsilon \to 0$ provided we 
keep $y=\epsilon\, N^{\psi}$ constant. For 1-XORSAT the exponent 
$\psi$ was found to be equal to $\frac 12$, and $\ln P_{SAT}$ to 
converge to the scaling function $\Phi _1(y)$ (\ref{sca1}). 
The situation is similar but slightly more
involved for 2-XORSAT. A natural assumption is to look for 
the existence of a scaling function  such that
\begin{equation}\label{sca2a}
\ln P_{SAT} ( N, \epsilon) \simeq  N^\rho \; 
\Phi _2 ( \epsilon \, N^\psi)  \ .
\end{equation}
Let us see if (\ref{sca2a}) is compatible with the limiting
behaviours (\ref{bornesp2}). Fixing $\epsilon<0$ and sending $N\to\infty$ 
we obtain, for $y=\epsilon N^\psi \to -\infty$,  
$\frac 14 \ln |y| - \frac \psi 4 \ln N$ for the l.h.s, and
$N^\rho \times \Phi_2(y)$ for the r.h.s. Hence $\rho=0$ as in the 
1-XORSAT case, but an additive correction is necessary, and we modify 
scaling Ansatz (\ref{sca2a}) into
\begin{equation}\label{sca2b}
\ln P_{SAT} ( N, \epsilon) \simeq  
\Phi _2 ( y= \epsilon \, N^\psi) - \frac{\psi}{4}\; \ln N .
\end{equation}
The above equation is now compatible with (\ref{bornesp2}) if
$\Phi_2(y) \sim \frac 14 \ln |y|$ when $y\to -\infty$.
Fixing now $\epsilon >0$ and sending $N$ to infinity we see that
(\ref{bornesp2}) is fulfilled if  $\Phi_2(y) \sim 
- \Omega\,  y^3$ when $y\to +\infty$ and 
\begin{equation} \label{valpsi}
\psi = \frac 13 \ .
\end{equation}
The above value for $\psi$ is expected from the study of random graphs
\cite{Bo89} and is related to the size 
$N^{1-\psi}=N^{\frac 23}$ of the largest components at the percolation
threshold (Section \ref{secrg}). 
$\psi$ is called 
critical exponent and characterize the width of the critical region 
of 2-XORSAT. Loosely speaking it means that a formula of with $N$
variables and $\frac N2+\Delta$ equations is 'critical'  when $\Delta \sim 
N^{\frac 23}$. This information will be useful for the analysis of
search algorithms in Section \ref{secuni}.

A consequence of (\ref{sca2b},\ref{valpsi}) is that, right at the
threshold, the probability of satisfaction decays as\footnote{This 
scaling is correct provided there is no diverging 
e.g. $O(\ln \ln N)$ corrections to (\ref{sca2b}).}
\begin{equation} \label{predic112}
P_{SAT}\left(N, \alpha_c \right) \sim N^{-\frac 1{12}} \ .
\end{equation}
This scaling agrees with numerical experiments, though the
small value of the decay exponent makes an accurate check delicate
(Inset of Figure \ref{proba2b-fig}).
  
\subsection{First and second moments inequalities for 
the 3-XORSAT threshold}

\begin{figure}[t]
\includegraphics*[width=5.truecm,angle=0]{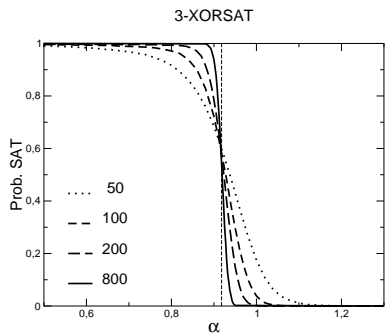}
\caption{Probability that a random 3-XORSAT formula is satisfiable as
a function of the ratio $\alpha$ of equations per variable, and for
various sizes $N$. The dotted line locates the threshold $\alpha_c
\simeq 0.918$.} 
\label{proba3-fig}
\end{figure}

Figure \ref{proba3-fig} shows the probability that a random 3-XORSAT
formula is satisfiable as a function of $\alpha$ for increasing sizes
$N$. It appears that formulas with ratio $\alpha < \alpha _c \simeq
0.92$ are very likely to be satisfiable in the large $N$ limit,
while formulas with ratios beyond this critical value are almost
surely unsatisfiable. This behaviour is different from the 2-XORSAT
case (Figure \ref{proba2-fig}) 
in that $P_{SAT}$ seems to tend to unity below threshold.

It is important to realize that, contrary to the 2-XORSAT case, the
Sat/Unsat transition is not related to connectivity percolation.
Consider indeed a variable, say, $x_1$. This variable appear, on
average, in $3\alpha$ equations. Each of those equations contain
other 2 variables. Hence the `connectivity' of $x_1$ is $c=6\alpha$,
which is larger than unity for $\alpha _p=\frac 16$. In the range
$[\alpha_p,\alpha_c]$ the formula is percolating but still satisfiable
with high probability. The reason is that cycles do not
hinder satisfiability as much as in the 2-XORSAT case.

Use of the first and second moment inequalities (\ref{appmoment}) for
the number ${\cal N}$ 
of solutions provides us with upper and lower bounds to the
Sat/Unsat ratio $\alpha_c$. The calculation follows the same line as 
the one of the 2-XORSAT case (Section \ref{secld2}). The
first moment $\langle {\cal N} \rangle =2 ^{N(1-\alpha)}$ vanishes for
ratios larger than unity, showing that
\begin{equation}
\alpha _c \le \alpha _1 =1 \ .
\end{equation}
This upper bound is definitely larger than the true threshold from the 
numerical findings of Figure~\ref{proba3-fig}. We have already
encountered
this situation in 2-XORSAT: in the $\frac 12 < \alpha < 1$ range 
formulas are unsatisfiable with probability one (when $N\to\infty$),
yet the average number of solutions is exponentially large! The reason
is, once more, that the average result is spoiled by rare, satisfiable
formulas with many solutions.

As for the second moment expression (\ref{n2p},\ref{upperom2}) still holds 
with $q(d)$ given by (\ref{qk}) with $K=3$. The absolute maximum of the
corresponding function $A(d,\alpha)$ is located in
$d^* = \frac 12$ when $\alpha < \alpha _2 \simeq 0.889$, and
$d^* < \frac 12$ when $\alpha > \alpha_2$. In the latter case 
$\langle {\cal N}^2\rangle$ is exponentially larger than $\langle
{\cal N}\rangle ^2$, and the second moment inequality
(\ref{ineq}) does not give any information about $P_{SAT}$. In
the former case  $\langle {\cal N}^2\rangle$ and
$\langle {\cal N}\rangle ^2$ are equivalent to exponential-in-$N$
order. It is shown in \ref{appfluctu} that their  ratio actually tends to 
one as $N\to\infty$. We conclude that formulas with
ratios of equations per variable 
less than $\alpha_2$ are satisfiable with high probability in
the infinite size limit, or, equivalently \cite{Cr03b},
\begin{equation}
\alpha _c \ge \alpha _2 \simeq 0.889 \ .
\end{equation}
Unfortunately the lower and upper bounds do not match and the precise
value of the threshold remains unknown at this stage. We explain in
the next section how a simple preprocessing of the formula, before the
application of the first and second moment inequalities, can close the
gap, and shed light on the structure of the space of solutions.

\subsection{Space  of solutions and clustering}\label{seccluster}

We start from a simple observation. Assume we have a formula $F$ 
of 3-XORSAT where a variable, say, $x$, appears only once, that is, in 
one equation, say, $E:x+y+z=0$. Let us call $F'$ the subformula obtained
from $F$ after removal of equation $E$. Then the following statement
is true: {\em $F$ is satisfiable if and only if $F'$ is satisfiable.}
The proof is obvious: whatever the values of $y,z$ required to satisfy
$F'$ equation $E$ can be satisfied by an adequate choice of $x$, and
so can be the whole formula $F$.

In a random 3-XORSAT formula $F$ with ratio $\alpha$ there are about
$N\times 3\alpha \, e^{-3\alpha}$ variables appearing only once in the
formula. Removal of those variables (and their equations) produces
a shorter formula with $O(N)$ less equations. Furthermore it may
happen that variables with multiple occurrences in the original
formula have disappeared from the output formula, or appear only
once. Hence the procedure can be iterated until no single-occurrence
variables are present. We are left with $F_2$, the largest
subformula (of the original formula) where every variable
appears at least twice.

Many questions can be asked: how many equations are left in
$F_2$? how many variables does it involve? how many solutions does it
have? Giving the answers requires a thorough 
analysis of the removal procedure, with the techniques exposed in Section
\ref{secuc} \cite{Co03,Me03,Du02}. 
The outcome depends on the value of the ratio 
compared to 
\begin{equation}
\alpha _d = \min _b -\frac{\log (1-b)}{3\, b^2} \simeq 0.8184\ldots
\end{equation}
hereafter called clustering threshold. With 
high probability when $N\to\infty$ $F_2$ is empty if $\alpha <
\alpha_d$, and contains an extensive number 
of equations, variables when $\alpha > \alpha_d$. In the latter case
calculation of the first and second moments 
of the number of solutions of $F_2$ shows that this number does
not fluctuate around the value $e^{N\,
  s_{cluster}(\alpha)+o(N)}$ where
\begin{equation}\label{scluster}
s_{cluster} (\alpha) = (b - 3 \alpha \,b^2 + 2 \alpha \, b^3)\ln 2
\end{equation}
and $b$ is the strictly positive solution of the self-consistent equation 
\begin{equation} \label{eqb}
1-b=e^{-3\, \alpha\, b^2} \ .
\end{equation}
Hence $F_2$ is satisfiable if and only if $\alpha < \alpha _c$ defined
through $s_{cluster} (\alpha_c)=0$, that is,
\begin{equation}
\alpha_c \simeq 0.9179 \ldots \ .
\end{equation}
This value is, by virtue of the equivalence between $F$ and $F_2$ 
the Sat/Unsat threshold for 3-XORSAT, in excellent agreement with 
Figure~\ref{proba3-fig}.

\begin{figure}
\begin{center}
\includegraphics[width=120pt,angle=0]{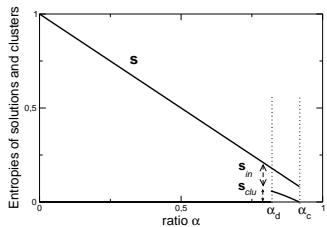}
\end{center}
\caption{Entropies (base 2 logarithms divided by
size $N$) of the numbers of solutions and clusters as a function of the 
ratio $\alpha$. The entropy of solutions  
equals $1-\alpha$ for $\alpha<\alpha_c\simeq 0.918$. 
For $\alpha<\alpha_d\simeq 0.818$, solutions are uniformly
scattered on the $N$-dimensional hypercube. 
At $\alpha_d$ the  solution space discontinuously
breaks into disjoint clusters. The entropies of clusters, $s_{cluster}$,
and of solutions in each cluster, $s_{in}$, are such that 
$s_{cluster}+s_{in}=s$. 
At $\alpha_c$ the number of clusters stops being exponentially large
($s_{cluster}=0$). Above $\alpha_c$ there is almost surely no solution.}
\label{entroxor}
\end{figure}

\begin{figure}
\begin{center}
\includegraphics[width=150pt]{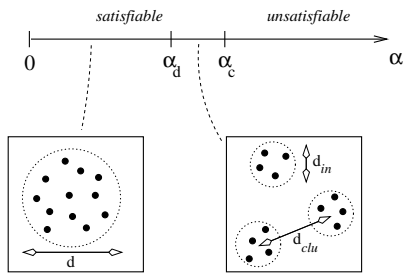}
\end{center}
\caption{Phase diagram of 3-XORSAT. A `geometrical' phase 
transition takes place in the satisfiable phase at  
$\alpha_d\simeq 0.818$. At small ratios $\alpha<\alpha_d$ 
solutions are uniformely
scattered on the $N$-dimensional hypercube, with a typical normalized 
Hamming distance $d=\frac 12$. At $\alpha_d$ the solution space discontinuously
breaks into disjoint clusters: the Hamming distance $d_{in}\simeq 0.14$ 
between solutions inside a cluster is much smaller than
the typical distance $d_{clu}=\frac 12$ between two clusters.}
\label{cluster-fig}
\end{figure}

How can we reconstruct the solutions of $F$ from the ones of $F_2$?
The procedure is simple. Start from one solution of $F_2$ (empty
string if $\alpha < \alpha_d$). Then introduce back the last  
equation which was removed since it contained $n\ge 1$ single-occurrence
variable. If $n=1$ we fix the value of this variable in a unique
way. If $n=2$ (respectively $n=3$) there are 2 (respectively, 4) ways
of assigning the reintroduced variables, defining as many solutions
from our initial, partial solution. Reintroduction of equations one after
the other according to the Last In -- First Out order gives us more and
more solutions from the initial one, until we get a bunch of solutions
of the original formula $F$. It turns out that the number of solutions
created this way is $e^{N\, s_{in}(\alpha)+o(N)}$ where
\begin{equation}\label{sin}
s_{in} (\alpha) = (1-\alpha)\,\ln 2 - s_{cluster}(\alpha) \ .
\end{equation}
The above formula is true for $\alpha> \alpha_d$, and should be
intended as $s_{in}(\alpha) =(1 - \alpha)\,\ln 2$ for  $\alpha< \alpha_d$.
These two entropies are shown in Figure \ref{entroxor}.
The total entropy, $s^*(\alpha) = s_{in}(\alpha)+s_{cluster}(\alpha)$,
is simply 
$(1-\alpha)\,\ln 2$ for all ratios smaller than the
Sat/Unsat threshold. It shows no singularity at the clustering threshold.
However a drastic change in the structure of the space of solutions
takes place, symbolized in the phase diagram of Figure \ref{cluster-fig}: 
\begin{itemize}
\item For ratios $\alpha<\alpha_d$ the intensive
Hamming distance between two solutions
is, with high probability, equal to $d=1/2$. Solutions thus differ 
on $N/2 +o(N)$ variables, as if they were statistically unrelated 
assignments of the $N$ Boolean variables. 
In addition the space of solutions enjoys some connectedness 
property. Any two solutions are connected by
a path (in the space of solutions) along which successive solutions
differ by a bounded number of variables. Losely speaking 
one is not forced to cross a big region prived of solutions
when going from one solution to another.
\item For ratios $\alpha > \alpha_d$ 
the space of solutions is not connected any longer.  It is made
of an exponentially large (in $N$) number ${\cal N}_{clu}
=e^{N\, s_{cluster}}$ of connected
components, called clusters, each containing an exponentially large
number ${\cal N}_{in}=e^{N\, s_{in}}$ of solutions.  Two solutions belonging to
different clusters lie apart at a Hamming distance $d_{clu}=1/2$ while,
inside a cluster, the distance is $d_{in} < d_{clu}$. $b$ given by
(\ref{eqb}) is the fraction of variables having the same value in all
the solutions of a cluster (defined as the backbone).
\end{itemize}

We present in Sections \ref{secreplicas} and \ref{seccavity} 
statistical physics
tools developed to deal with the scenario of Figure \ref{cluster-fig}.

\section{Advanced methods (I): replicas} \label{secreplicas}

\subsection{From moments to large deviations for the entropy}
\label{secmoment}

The analysis of Section \ref{secld2} has shown that the first, and second
moments of the number ${\cal N}$ of solutions
are dominated by rare formulas with a lot of solutions. 
Let us define the intensive entropy $s$ through ${\cal N} = e^{N\,
s}$. As ${\cal N}$ is random (at fixed $\alpha,N$) so is $s$. We
assume that the distribution of $s$ can be described, in the large
size limit, by a rate function $\omega(s)$ (which depends on $\alpha$).  
Hence,
\begin{equation}
\langle {\cal N}^q \rangle = \int ds\; e^{-N \, \omega(s)} \times
\left(e^{N\, s}\right)^q \sim \exp\big[N  \max _{s} \big(q\, s - 
\omega (s) \big)\big] 
\end{equation}
using the Laplace method. If we are
able to estimate the leading behaviour of the $q^{th}$ moment of
the number of solutions when $N$ gets large at fixed $\alpha$,
\begin{equation}
\langle {\cal N}^q \rangle \sim e^{N\, g(q)} \ ,
\end{equation}
 then $\omega$ can be easily
calculated by taking the Legendre transform of $g$. In particular
the typical entropy is obtained by $s^* = \frac{dg}{dq}(q\to 0)$.
This is the road we will follow below. We will show how $g(q)$ can be 
calculated when $q$ takes integer values, and then perform an analytic
continuation to non integer $q$. The continuation leads to substantial 
mathematical difficulties, but is not uncommon in statistical physics
e.g. the $q\to 1$ limit of the $q$-state Potts model to recover
percolation, or
the $n\to 0$ limit of the $O(n)$ model to describe self-avoiding walks.

To calculate the $q^{th}$ moment we will have to average over the
random components of formulas $F$, that is, the $K$-uplets of index
variables in the first members and the $v=0,1$ second
members. Consider now  homogeneous formulas $F_h$ whose first
members are randomly drawn in the same way as for $F$, but with all
second members $v=0$. The number ${\cal N}_h$ of solutions of a 
homogeneous formula is always larger or equal to one. It is a simple
exercise to show that
\begin{equation} \label{momentxorequality}
\langle {\cal N} ^{q+1} \rangle =
2^{N(1-\alpha)}\times \langle \big({\cal N}_h\big)^q \rangle \quad ,
\end{equation}
valid for any positive integer $q$\footnote{Actually the identity
  holds for $q=0$ too, and is known under the name of harmonic
mean formula \cite{Al89}.}. Therefore it is sufficient to calculate
the moments of ${\cal N}_h=e^{N\, g_h(q)}$  since 
(\ref{momentxorequality}) gives a simple identity between
$g(q+1)$ and $g_h(q)$. This technical simplification has a
deep physical meaning  we will  comment in Section \ref{secpin}. 

\subsection{Free energy  for replicated variables}

The $q^{th}$ power of the number of solutions to a homogeneous system
reads
\begin{equation}\label{qxor1}
\big({\cal N}_h\big)^q = \left[ \sum _X \prod _{\ell=1}^M 
e_\ell ( X) \right] ^q = \sum _{X^1,X^2, \ldots , X^q} 
\prod _{\ell=1}^M \prod_{a=1}^q e_\ell ( X^a) \ ,
\end{equation} 
where $e_\ell(X)$ is 1 if equation $\ell$ is satisfied by assignment
$X$. The last sum runs over $q$ assignments $X^a$, with $a=1,2, \ldots ,
q$ of the Boolean variables, called replicas of the original
assignment $X$.
It will turn useful to denote by $\vec x_i = (x_i^1,x_i^2, \ldots ,
x_i^q)$ the $q$-dimensional vector whose components are the values of
variable $x_i$ in the $q$ replicas. To simplify notations we consider
the case $K=3$ only here, but extension to other values of $K$ is
straightforward. 
Averaging over the instance, that is, the triplets of integers labelling
the variables involved in each equation $\ell$, leads to the following
expression for the $q^{th}$ moment, 
\begin{eqnarray} \label{qxor2}
\langle\big({\cal N}_h\big)^q \rangle&=& \sum _{X^1,X^2, \ldots , X^q} 
\langle \prod_{a=1}^q e ( X^a) \rangle ^M \nonumber \\
&=& \sum _{X^1,X^2, \ldots , X^q} \left[ \frac 1{N^3}
\sum _{1\le i,j,k\le N} \delta _{\vec x_i + \vec x_j + \vec x_k }
+ O\bigg(\frac 1N\bigg)\right] ^M
\end{eqnarray} 
where $\delta _{\vec x} =1$ if the compoments of $\vec x$ are all
null mod. 2, and 0 otherwise.
We now procede to some formal manipulations of the above equation
(\ref{qxor2}). 

{\bf First step.}
Be ${\cal X} = \{X^1,X^2,\ldots , X^q\}$ one of the $2^{\,qN}$ replica 
assignment. Focus on variable $i$, and its attached assignment
vector, $\vec x_i$. The latter may be any of the $2^q$ possible vectors 
 e.g. $\vec x_i = (1,0,1,0,0, \ldots , 0)$ if variable $x_i$ is equal
to $0$ in all but the first and third replicas. The histogram of 
the assignments vectors given replica assignment ${\cal X}$,  
\begin{equation} \label{qxor2b}
\rho \big( \vec x |{\cal X}\big) = 
\frac 1N \sum _{i=1}^N \delta _{\vec x- \vec x_i} 
\quad ,
\end{equation}
counts the fraction of assignments vectors $\vec x _i$ having value
$\vec x$ when $i$ scans the whole set of variables from 1 to $N$.
Of course, this histogram is normalised to unity,
\begin{equation}\label{qxor2c}
\sum _{\vec x} \rho \big( \vec x\big) = 1
\quad ,
\end{equation}
where the sum runs over all $2^q$ assignment vectors.
An simple but essential observation is that the r.h.s. of
(\ref{qxor2}) may be rewritten in terms of the above histogram,
\begin{equation} \label{qxor3}
\frac 1{N^3}
\sum _{1\le i,j,k\le N} \delta _{ \vec x_i + \vec x_j + \vec x_k }
= \sum _{\vec x , \vec x'} 
\rho \big( \vec x \big) \;\rho \big( \vec x'\big) \;\rho \big( \vec 
x + \vec x' \big) 
\ .
\end{equation} 
Keep in mind that $\rho$ in (\ref{qxor2b},\ref{qxor3}) depends on the
replica assignement ${\cal X}$ under consideration.\\

{\bf Second step.} According to (\ref{qxor3}), two
replica assignments ${\cal X}_1$ and ${\cal X}_2$ defining the same
histogram $\rho$ will give equal contributions to 
$\langle\big({\cal N}_h\big)^q\rangle$.
The sum over replica assignments ${\cal X}$ can therefore be replaced
over the sum over possible histograms provided the multiplicity ${\cal
M}$ of the latter is taken properly into account.
This multiplicity is also equal to the number of combinations
of $N$ elements (the $\vec x_i$ vectors) into $2^q$ sets labelled
by $\vec x$ and of cardinalities $N\, \rho(\vec x)$. We obtain
\begin{equation}\label{qxor5}
\langle \big({\cal N}_h\big)^q \rangle= \sum _{\{ \rho\}} ^{(norm)}
e^{\displaystyle{\;  N \; {\cal G}_h \big( \{\rho\},\alpha \big) + o(N) }}
\quad ,
\end{equation} 
where the $(norm)$ subscript indicates that the sum runs
over histograms $\rho$ normalized according to (\ref{qxor2c}), and
\begin{equation}\label{qxor6}
{\cal G}_h \big( \{\rho\},\alpha \big) = - \sum _x \rho(x) \, \ln \rho(x) +
\alpha\; \ln \bigg[ \sum _{\vec x , \vec x'} 
\rho \big( \vec x \big) \;\rho \big( \vec x'\big) \;
\rho \big( \vec x +\vec x' \big) 
\bigg]
\ .
\end{equation} 
In the large $N$ limit, the sum in (\ref{qxor5}) is 
dominated by the histogram $\rho ^*$ maximizing the functional ${\cal G}_h$.

{\bf Third step.}
 Maximisation of function ${\cal G}_h$
over normalized histograms can be done within the Lagrange multiplier 
formalism. The procedure consists in considering the modified function  
\begin{equation}
{\cal G}^{LM}_h \big( \{\rho\} , \lambda , \alpha \big) =
{\cal G}_h \big( \{\rho\} , \alpha \big) + \lambda \; \bigg( 1 -
\sum _{\vec x} \rho \big( \vec x\big)  \bigg) \quad ,
\end{equation}
and first maximise ${\cal G}^{LM}_h$ with respect to
histograms $\rho$ without caring about the normalisation constraint,
and then optimise the result with respect to $\lambda$.
We follow this procedure with ${\cal G}_h$ given by (\ref{qxor6}). 
Requiring that ${\cal G}^{LM}_h$ be maximal provides us with a set
of $2^q$ coupled equations for $\rho^*$, 
\begin{equation} \label{spxor}
\ln \rho ^*(\vec x ) + 1 +\lambda -3\;\alpha\ \frac
{\displaystyle{ \sum _{\vec x' } 
\rho ^* \big( \vec x'\big) \;
\rho ^*\big( \vec x+\vec x' \big) }}
{\displaystyle{ \sum _{\vec x' , \vec x''} 
\rho ^*\big( \vec x' \big) \;\rho ^*\big( \vec x''\big) \;
\rho ^*\big( \vec x'+\vec x '' \big) }}
= 0 \ ,
\end{equation}
one for each assignment vector $\vec x$. The optimisation equation 
over $\lambda$ implies that $\lambda$ in  (\ref{spxor}) is such that 
$\rho^*$ is normalised.
At this point of the above and rather abstract calculation it may help
to understand the interpretation of 
the optimal histogram $\rho^*$. 

\subsection{The order parameter}

We have already addressed a similar question at the end 
of the second moment calculation in Section \ref{secop}. 
The parameter $d^*$ coming out from
the calculation was the (weighted) average Hamming distance 
(\ref{davstar}) between two solutions of the same random instance. 
The significance of $\rho^*$ is identical. Consider $q'$ solutions
labelled by $a=1,2,\ldots, q'$ of the same random and homogeneous 
instance and a variable, say, $x_i$. What is the probability,
over instances and solutions, that this variable takes, for instance,
value 0 in the first and fourth solutions, and 1 in all other solutions?
In other words, what is the probability that the assignment vector 
$\vec x _i = (x_i^1, x_i^2, \ldots , x_i^{q'})$ 
is equal to $\vec x' = (0,1,1,0,1,1,\ldots,1)$? The answer is
\begin{equation}\label{defpxor}
p(\vec x') = \left\langle \frac 1{({\cal N}_h)^{q'}} 
\sum _{X^1,X^2, \ldots, X^{q'}}\;\delta _{\vec x_i - \vec x} \;
\prod _{l=1}^M \prod _{a=1}^q e_\ell ( X^a) \right\rangle
\end{equation}
where the dependence on $i$ is wiped out by the average over the instance. 
The above probability is an interesting quantity; it provides us information
about the `microscopic' nature of solutions. Setting $q'=1$ gives
us the probabilities $p(0),p(1)$ that a variable is false or true
respectively, that is, takes the same value as in the null assignment or not.
For generic $q'$ we may think of two extreme situations:
\begin{itemize}
\item a flat $p$ over assignment vectors, $p(\vec x ')=1/2^{q'}$, corresponds
to essentially orthogonal solutions;
\item on the opposite, a concentrated probability e.g. $p(\vec x ')=
\delta_{\vec x'}$ implies that variables
are extremely constrained, and that the (almost) unique solution
is the null assignment.
\end{itemize}

The careful reader will have already guessed that our calculation
of the $q^{th}$ moment gives access to a weighted counterpart of $p$.
The order parameter 
\begin{equation} \label{defrhoetoilexor}
\rho ^*(\vec x) = \frac 1{\langle ({\cal N}_h)^q\rangle}
\sum _{X^1,X^2, \ldots, X^q}\; \delta _{\vec x_i - \vec x} \;
\left\langle \prod _{l=1}^M \prod_{a=1}^q  e_\ell ( X^a) \right\rangle \quad ,
\end{equation}
is not equal to $p$ 
even when $q=q'$. However, at the price
of mathematical rigor, the exact probability $p$ over 
vector assignments of integer length $q'$ can be reconstructed from
the optimal histogram $\rho ^*$ associated to moments of
order $q$ when $q$ is real-valued and sent to $0$. 
The underlying idea is the following. Consider
(\ref{defrhoetoilexor}) and an integer $q'<q$. From any assignment
vector $\vec x$ of length $q$, we define two assignment vectors 
$\vec x', \vec x''$ of respective lengths $q', q-q'$ corresponding to
the first $q'$ and the last $q-q'$ components of $\vec x$ respectively. 
Summing (\ref{defrhoetoilexor}) over the $2^{q-q'}$ assignment
vectors $\vec x''$ gives,
\begin{equation} \label{defrhoetoilexor2}
\sum _{\vec x''} \rho ^*(\vec x', \vec x'') 
= \frac 1{\langle ({\cal N}_h)^q \rangle}
\sum _{\{X^a\}}\; \delta _{\vec x'_i - \vec x'} \;
\left\langle \big({\cal N}_h\big) ^{q-q'}
\prod _{l,a} e_\ell ( X^a) \right\rangle \ .
\end{equation}
As $q$ now appears in the powers of ${\cal N}_h$ in the numerator and
denominator only, it can be formally send to zero at fixed $q'$, 
yielding
\begin{equation} \label{replicaess}
\lim _{q\to 0}\ \sum _{\vec x''} \rho ^*(\vec x', \vec x'') = p(\vec x ')
\end{equation}
from (\ref{defpxor}).
This identity justifies the denomination order parameter given to
$\rho^*$. 

Having understood the significance of $\rho^*$ helps us to find
appropriate solutions to (\ref{spxor}). Intuitively and from
the discussion of the first moment case $q=1$,
$p$ is expected to reflect both the special role of the null assignment
(which is a solution to all homogeneous systems) and the ability of
other solutions of a random system to be essentially orthogonal to 
this special assignment. A possible guess is thus
\begin{equation} \label{rhoetoilexor2}
p(\vec x ') = \frac {1-b}{2^{q'}}+ b \; \delta 
_{\vec x'} \quad ,
\end{equation}
where $b$ expresses some degree of `correlation' of solutions with
the null one. Hypothesis (\ref{rhoetoilexor2}) interpolates between the 
fully concentrated ($b=1$) and flat ($b=0$) probabilities.
$b$ measures the fraction of variables (among the $N$ ones) 
that take the 0 values in all $q'$ solution, and coincides with the
notion of backbone introduced in Section \ref{seccluster}. Hypothesis 
(\ref{rhoetoilexor2}) is equivalent, from the connection (\ref{replicaess})  
between $p$ and the annealed histogram $\rho ^*$ to the following
guess for the solution of the maximisation condition (\ref{spxor}),
\begin{equation} \label{rhoetoilexor}
\rho ^*(\vec x ) = \frac {1-b}{2^q}+ b \; \delta
_{\vec x} \ .
\end{equation}
Insertion of Ansatz (\ref{rhoetoilexor}) in (\ref{spxor}) shows
that it is indeed a solution provided $b$ is shrewdly chosen
as a function of $q$ and $\alpha$, $b=b^*(q,\alpha)$. 
Its value can be either found from direct resolution of (\ref{spxor}),
 or from insertion of histogram (\ref{rhoetoilexor}) 
in ${\cal G}_h$ (\ref{qxor6}) and maximisation over $b$, with the
result, 
\begin{equation} \label{xormaxgh}
g_h(q,\alpha) =  \max _{0\le b\le 1} A_h (b,q,\alpha)
\end{equation}
where
\begin{eqnarray} \label{xormaxgh2}
A _h (b,q,\alpha) &=& -\left( 1 -\frac 1{2^q} \right)
\; (1-b) \; \ln \left(\frac{1-b}{2^q}\right) \\
&-& \left( b + \frac{1-b}{2^q} \right) \; \ln 
\left( b + \frac{1-b}{2^q} \right) + \alpha\; 
\ln \left( b^3 + \frac{1-b^3}{2^q} \right) \ ,  
\nonumber 
\end{eqnarray}
where the maximum is precisely reached in $b^*$. Notice that, since
$\rho^*$ in (\ref{rhoetoilexor}) is entirely known from the value
of $b^*$, we shall indifferently call order parameter $\rho^*$, or
$b^*$ itself.

\subsection{Results}\label{secresu}

\begin{figure}
\begin{center}
\includegraphics[width=110pt,angle=-90]{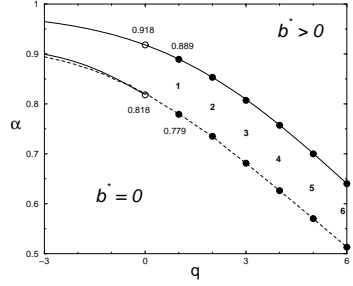}
\end{center}
\caption{The $q,\alpha$ 
plane and the critical lines $\alpha_M(q)$ (dashed),
$\alpha^*(q)$ (full tick), and $\alpha_s(q)$ (full thin) 
appearing in the calculation
of the $q^{th}$ moment for homogeneous 3-XORSAT systems. 
Full dots correspond to integer $q$ values, while
continuous curves result from the analytic continuation to
real $q$. The fraction of variables in the backbone, $b^*$, vanishes below the 
line $\alpha_M(q)$; the global maximum of $A_h$ in 
(\ref{xormaxgh2}) is located in $b^*>0$ for ratios $\alpha>\alpha^*(q)$.
Ansatz (\ref{rhoetoilexor}) is locally unstable in the hardly visible 
domain $q<0,\alpha_M(q) <\alpha <\alpha _s(q)$.}
\label{qetoilexor}
\end{figure}

Numerical investigation of  $A _h$ (\ref{xormaxgh2}) shows
that: for $\alpha<\alpha_M(q)$ the only local maximum of $A _h$ 
is located in $b^*=0$, and $A_h(q,\alpha)=q(1-\alpha)\ln 2$; 
when $\alpha_M(q) < \alpha<\alpha^*(q)$, there exists another local maximum 
in $b>0$ but the global maximum is still reached in $b^*=0$;
when $\alpha>\alpha^*(q)$, the global maximum is located in $b^*>0$.
This scenario extends to generic $q$ the findings of the second moment
calculation carried out in Section
\ref{secld2}. The $\alpha_M$ and $\alpha^*$ lines divide the
$q,\alpha$  plane as
shown in Figure~\ref{qetoilexor}. Notice that, while the black dots in
Figure~\ref{qetoilexor} correspond to integer-valued $q$, the continuous
lines are the output of the implicit analytic continuation to real $q$
done by the replica calculation.  

\begin{figure}
\begin{center}
\includegraphics[width=120pt,angle=0]{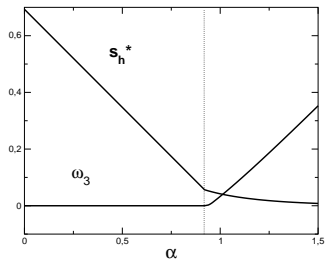}
\end{center}
\caption{Rate function $\omega _3$ for the probability of satisfaction
of full (bottom curve) and entropy $s_h^*$ of solutions
for homogeneous (top curve) 3-XORSAT systems vs. $\alpha$. 
The vertical dotted lines indicate the critical
Sat/Unsat threshold, $\alpha_c\simeq 0.918$. For $\alpha <\alpha_c$
$\omega _3=0$, and  $s_h^*=(1-\alpha)\,\ln 2$ is the same as for full systems.
Above the threshold  $\omega ^* <0$. Homogeneous
systems are, of course, always satisfiable: the entropy $s_h^*$ is a positive
but quickly decreasing function of $\alpha$.}
\label{xorcurve}
\end{figure}

Taking the derivative of (\ref{xormaxgh}) with respect to $q$ and
sending $q\to 0$ we obtain the typical entropy of a homogeneous
3-XORSAT system at ratio $\alpha$,
\begin{equation}\label{commentxor1}
s_h ^*(\alpha)=  \ln 2 \times \max _{0\le b\le 1} 
\big[ (1-b) \big(1 - \ln(1-b)\big) - \alpha \, (1-b^3) \big] \ . 
\end{equation} 
The optimal value for $b$ coincides with the solution of (\ref{eqb}). 
The typical entropy is plotted in Figure~\ref{xorcurve}, and is equal to:
\begin{itemize}
\item $(1-\alpha) \ln 2$ when $\alpha<\alpha_c\simeq 0.918$ 
(Figure~\ref{qetoilexor}); in this range of ratios, homogeneous and
  full (with random second members)
systems have essentially the same properties, with the same cluster
 organisation of solutions, and identical entropies of solutions. 
\item a positive but rapidly decreasing function given by
(\ref{commentxor1}) when $\alpha>\alpha_c$; above the
critical ratio, a full system has no solution any more, while a homogeneous
instance still enjoys a positive entropy. The expression for
$s^*_h(\alpha)$ coincides with the continuation to $\alpha>\alpha_c$ of 
the entropy $s_{in}(\alpha)$ (\ref{sin}) of solutions in
a single cluster for a full system.
In other words, a single cluster of solutions, the one with the 
null solution, survive for ratios $\alpha>\alpha_S$ in homogeneous systems.  
\end{itemize}

Atypical instances can be studied and the large deviation rate 
function for the entropy can be derived from (\ref{xormaxgh}) for
homogeneous systems, and using equivalence (\ref{momentxorequality}), 
for full systems. Minimizing over the entropy we obtain the rate
function $\omega_3(\alpha)$ associated to the probability that a
random 3-XORSAT system is satisfiable,
with the result shown in Figure~\ref{xorcurve}. As expected we find
$\omega_3 =0$ for $\alpha < \alpha_c$ and $\omega_3 >0 $ for
$\alpha>\alpha_c$, allowing us to locate the Sat/Unsat threshold.

Notice that the emergence of clustering can be guessed from Figure 
\ref{qetoilexor}. It coincides with the appearance of a local
maximum of $A_h$ (\ref{xormaxgh2}) with a non vanishing 
backbone $b$. While in the intermediate phase $\alpha_d<\alpha<
\alpha_c$, the height of the 
global maximum equals the total entropy $s^*$, the height of the local 
maximum coincides with the entropy of clusters $s _{cluster}$ 
(\ref{scluster}).

\subsection{Stability of the replica Ansatz}

The above results rely on Ansatz (\ref{rhoetoilexor}). A necessary criterion
for its validity is that $\rho^*$ locates a true local maximum
of ${\cal G}_h$ , and not merely a saddle-point. Hence we have to calculate
the Hessian  matrix of ${\cal G}_h$ in $\rho^*$, and check that the eigenvalues
are all negative \cite{De79}. 
Differentiating (\ref{qxor6}) with respect to $\rho(\vec x)$
and $\rho (\vec x')$ we obtain the Hessian matrix
\begin{equation} \label{hvalue}
H(\vec x,\vec x') = -\frac{\delta _{\vec x+\vec x'}}{\rho^* (\vec x)}
+6\alpha\; \frac{\rho^* (\vec x+\vec x')}D - 9\alpha \;\frac{N(\vec x)}D
  \, \frac{N(\vec x')}D\ ,
\end{equation}
where $D=\frac {1-b^3}{2^q} + b^3$, $N(\vec x) = \frac {1-b^2}{2^q} 
+ b^2\,\delta _{\vec x}$. We use $b$ instead of $b^*$ to 
ligthen the notations, but it is intended that $b$ is the backbone
value which maximizes $A_h$ (\ref{xormaxgh2}) at fixed $q,\alpha$.
To take into account the global constraint over the histogram 
(\ref{qxor2c}) one can express one fraction, say, $\rho (\vec 0)$, as
a function of the other fractions $\rho(\vec x)$, $\vec x\ne \vec 0$.
${\cal G}_H$ is now a fonction of $2^q-1$ independent variables, 
with a Hessian matrix $\tilde H$ simply related to $H$,
\begin{equation} \label{htilde}
\tilde H(\vec x,\vec x') = H(\vec x,\vec x') - H(\vec x,\vec 0) -
H(\vec 0,\vec x')  + H(\vec 0,\vec 0) \ .
\end{equation}
Plugging expression (\ref{hvalue}) into (\ref{htilde}) we obtain  
\begin{eqnarray}
\tilde H(\vec x,\vec x') &=& \lambda _R\; \delta_{\vec x+\vec x'}
+ \frac 1{2^q-1}\big( \lambda _L - \lambda _R) \qquad \mbox{where} 
\nonumber \\
\lambda_R &=& 6\alpha \; \frac bD - \frac{2^q}{1-b} \\
\lambda _L &=& 2^q \left( 6\alpha \; \frac bD - \frac{2^q}
{(1-b)(1-b+2^qb)} - 9\alpha (1-2^{-q})\frac{b^4}{D^2}\right) \ .
\nonumber 
\end{eqnarray}
Diagonalization of $\tilde H$ is immediate, and we find two
eigenvalues:
\begin{itemize}
\item $\lambda _L$ (non degenerate). The eigenmode corresponds
to a uniform infinitesimal
variation of $\rho (\vec x$) for all $\vec x\ne \vec 0$, that is, a
change of $b$ in (\ref{rhoetoilexor}). It is an easy check that 
\begin{equation}
\lambda _L = \frac{2^q}{1-2^{-q}}\; \frac{\partial ^2 A_h}{\partial
b^2} (b,q,\alpha) \ ,
\end{equation}
where $A_h$ is defined in (\ref{xormaxgh2}).
As we have chosen $b$ to maximize $A_h$ this mode, called longitudinal 
in replica literature \cite{De79}, is stable\footnote{Actually $b^*$
is chosen to {\em minimize} $A_h$ when $q<0$, thus $\lambda_L$ has
always the right negative sign.}. 
 
\item $\lambda _R$ ($2^q-2$-fold degenerate): the eigenmodes correspond
to fluctuations of the order parameter $\rho$ transverse to
the replica subspace described by (\ref{rhoetoilexor}), and are
called replicon in spin-glass theory\cite{De79}. Inspection of 
$\lambda_R$ as a function
of $\alpha,q$ shows that it is always negative when $q>0$. For $q<0$ 
the replicon  mode is stable if 
\begin{equation}
\alpha > \alpha_s (q) = \frac{1- b^3 + 2^q\, b^3}{6 \, b(1-b)} \ .
\end{equation}
which is a function of $q$ only once we have chosen $b=b^*(q,\alpha_s)$.
\end{itemize}
The unstable region $q<0, \alpha_M (q) < \alpha <\alpha _s(q)$ is 
shown in Figure \ref{qetoilexor} and is hardly visible when $q>-3$. 
In this region a continuous symmetry breaking is expected \cite{Me87}.
In particular $\alpha_s$ stay below the $\alpha^*$ line for
small (in absolute value) and negative $q$. We conclude 
that our Ansatz (\ref{rhoetoilexor}) defines a maximum
of ${\cal G}_h$. 

Is it the global maximum of ${\cal G}_h$? There is no simple way to 
answer this question. Local stability does not rule out the possibility
for a discontinuous transition to another maximum in the replica order
parameter space not described by (\ref{rhoetoilexor}). 
A final remark is that a similar calculation can be done for any value
of $K$. The outcome for $K=2$ is the rate function $\omega_2$ plotted
in Figure~\ref{proba2b-fig}, in good agreement with numerics
close to the threshold.

\section{Advanced methods (II): cavity}\label{seccavity}

The cavity method, in the context of disordered systems, was 
historically developed as an alternative to the the replica method
\cite{Me87}. Its application to spin systems on random graphs is
extensively explained in \cite{Me01y}, and we limit ourselves here 
to briefly show how it gives back the 3-XORSAT scenario of Section
\ref{seccluster} \cite{Me03}. 

Let us consider a system $F$ involving variables $x_i$, $i=1,\ldots,
N$. In the following we will indifferently 
use the variable $x_i=0,1$ or its spin  representation
$S_i=(-1)^{x_i}=\pm 1$ when convenient. 
Let us define the GS energy $E^F _i (S_i)$ 
of the system when the $i^{th}$ spin is kept fixed, that is, the
minimal number of violated equations in $F$, taken over the $2^{N-1}$ 
configurations. We may always write 
\begin{equation} \label{defec}
E^F_i (S_i) = -w_i-h_i\, S_i\ ,
\end{equation}
where $h_i$ is called `field' acting on spin $S_i$.
For a homogeneous system $E^F_i(+1)=0$, and $E^F_i(-1)=n_i$ for some integer
$n_i$. Hence $h_i=\frac {n_i}2$ takes half-integer values.

The above definition can be  extended to the case of
$\ell > 1$ fixed spins. Let $I\subset \{1,2,\ldots, N\}$ be a subset
of the indices of cardinal $|I|\ge 2$, and $S_I$ denote one of the $2^{|I|}$
configurations of the spins $S_i, i\in I$. The GS energy of $F$ for given 
$S_I$  can in general be written as   
\begin{equation} \label{defec2}
E^F_{I} (S_I) = -w_{I} - \sum _{i \in I}  h_{i} \, S_{i}
-\sum  _{I'\subset I : |I'|\ge 2} J_{I'}\,\prod
_{i\in I' }  S_i
\end{equation}
where the $h_i$s are the fields and the $J_{I'}$s are
effective couplings between subsets of spins. 

The basic cavity assumption is that effective couplings are
vanishingly small: $J_{I'}=0$ for every subset $I'$. This apparently bold
hypothesis critically relies on a general property of random graphs
(from which our system is built on). Define the distance between two
vertices as the minimal number of edges on pathes linking these two
points. Then vertices in a finite subset are, with high probability
when $N\to\infty$, typically at infinite distance from each
other\footnote{An alternative formulation is, for finite size $N$,
  that the shortest loops (in extensive number) have lengths of the order of
  $\log N$ \cite{Bo89}.}. When correlations 
between variables in GS extinguish with the distance {\em i.e.} when
the correlation length is finite the cavity assumption is
correct in the large $N$ limit  \cite{Me87,Mo05}. The assumption will break 
down when correlations subsist on infinite distance, which 
happens to be the case in the clustered phase. 

\subsection{Self-consistent equation for the fields} \label{seccav1}

Under the assumption that couplings between randomly picked up 
spins are null we are left with the fields only. The goal of this
section is to show how to calculate those fields, or more precisely,
their probability distribution. The derivation is based on the 
addition procedure 
already used in the calculation of the size of the giant component in
random graphs (Section \ref{secrg}). 

Consider a system $F$ over $N$ variables to which we want
to add one 
equation involving one new variable $S$, and two variables $S_1,S_2$
appearing in $F$. The energy function 
associated to this equation is
\begin{equation}
e(S,S_1,S_2) = \frac 12 \big( 1 - \sigma\, S\, S_1\, S_2 \big)
\end{equation}
where $\sigma=+1$, respectively $-1$, when the second member of the 
equation is 0, resp. 1. Let us calculate the GS
energy of the new system $F'=F$ +  added equation when the new
variable $S$ is kept fixed, 
\begin{equation}
E^{F'} (S) = \min _{S_1,S_2}\big[ e(S,S_1,S_2) + E^F_{1,2}
  (S_1,S_2) \big] = - w  - u \, S \ .
\end{equation}
With the cavity hypothesis the couplings between spins $S_1,S_2$ is
null and the minimization is straightforward. We deduce the 
following explicit
expression for the field acting on $S$ (called bias in the cavity
literature \cite{Me01y}), 
\begin{equation}\label{expressu}
u=\frac \sigma2\; \mbox{sign} \big(h_1\; h_2\big) \ .
\end{equation}
Suppose we now add $\ell \ge 1$ (and not only one) equations. The above
calculation can be easily repeated. The absence of couplings make the
total field acting on $S$ a linear combination of the fields coming
from each new equation, 
\begin{equation}\label{uj}
h = \sum _{j=1}^\ell u ^{j} \ ,
\end{equation}
where $u^{j}$ is calculated from (\ref{expressu}) and each pair of fields
$(h_1^{j}, h_2^{j})$ acting on the spins in the $j^{th}$ equation,
$j=1,\ldots, \ell$. 

How many equations should we add for our new system over $N+1$
variables to have the same statistical features as old one over $N$ 
variables? First $\ell$ should be Poisson distributed 
with parameter $3\alpha$. Then, given $\ell$, we randomly chose $\ell$
pairs of variables; for each pair the corresponding bias
$u$ can be calculated from (\ref{expressu}). Assume the output is a
set of $\ell$ independent biases, taking values
\begin{equation}\label{uk}
u = \left\{ \begin{array}{c c c}
+ \frac 12 & \mbox{with probability} & a_+ \\
0 & \mbox{with probability} & a_0 \\
- \frac 12 & \mbox{with probability} & a_- 
\end{array} \right. \ 
\end{equation}
Obviously $a_++a_0+a_-=1$.
Summing over the equations as in (\ref{uj}) we obtain the distribution of the
field $h$ acting on the new spin at fixed $\ell$,
\begin{equation}
p(h | \ell )=\sum _{\ell_+,\ell_0,\ell_-} {\ell \choose
  \ell+,\ell_0,\ell_-}
a_+^{\ell_+}\, a_0^{\ell_0}\, a_-^{\ell_-}\, \delta
  _{h-\frac 12(\ell_+-\ell_-)} \ .
\end{equation} 
Finally we sum over the Poisson distribution for $\ell$ to obtain the
distribution of fields $h$,
\begin{equation} \label{defph}
p(h)=e^{-3\alpha (1-a_0)} \sum _{\ell_+,\ell_-} \frac{
  (3\alpha)^{\ell_++\ell _-}}{\ell_+!\ell_-!} a_+^{\ell_+}\, 
 a_-^{\ell_-}\, \delta
  _{h-\frac 12(\ell_+-\ell_-)} \ .
\end{equation}
In turn we calculate the distribution of the biases from the one of
the fields through (\ref{expressu}). The outcome are the values of the
probabilities (\ref{uk}) in terms of $p$, 
\begin{eqnarray}
a_+ &=& \sum _{h_1,h_2: h_1 h_2 > 0} p(h_1)\, p(h_2) \ , 
\quad 
a_- = \sum _{h_1,h_2: h_1 h_2 < 0} p(h_1)\, p(h_2) \ ,\nonumber \\
a_0 &=& \sum _{h_1,h_2: h_1 h_2 = 0} p(h_1)\, p(h_2) = 2\, p(0) -p(0)^2 \ .
\label{conda0} 
\end{eqnarray}
The above equations together with (\ref{defph}) define three
self-consistent conditions for $a_0,a_+,a_-$. Notice that the free
energy can be calculated along the same lines \cite{Me01y}.

\subsection{Application to homogeneous and full systems}\label{seccavapp}

In the case of homogeneous systems ($\sigma = +1$) 
we expect all the fields to be
positive, and look for a solution of (\ref{conda0}) with $a_-=0$. Then
$p(h)$ (\ref{defph})
is a Poisson distribution for the integer-valued variable $2h$,
with parameter $3\,\alpha\, a_+$. 
The self-consistent equation (\ref{conda0}) reads
\begin{equation}
a_0=1-a_+= 2 \; e^{-3 \alpha a_+} - e^{-6\, \alpha a_+} = 1 -
\big( 1 -  e^{-3\, \alpha a_+} \big)^2
\end{equation}
which coincides with (\ref{eqb}) with the definition $b=\sqrt{a_+}$.
As expected
\begin{equation}
b= \sum _{h\ge \frac 12} p(h)  
\end{equation}
is the fraction of frozen variables (which cannot be flipped from 0 to
1 in GS assignments), in agreement with the notion of backbone of Section
\ref{seccluster}.

The energy is zero at all ratio $\alpha$ by construction. As for the
entropy consider adding a new equation to the system $F$ (but with no
new variable). With probability
$1-b^3$ at least one of the three variables in the new equation $e$ was
not frozen prior to addition, and the number of solutions of the new 
system $F+e$ is half the one of $F$. With probability $b^3$ all three
variables are frozen in $F$ (to the zero value) and the number of
solutions of $F+e$ is the same as the one of $F$. Hence the average
decrease in entropy is
\begin{equation} \label{evolsh}
N\, s_h^* (\alpha + \frac 1N) - N \, s_h^*(\alpha) \simeq 
\frac{ds ^*_h}{d\alpha} = -(1-b^3) \; \ln 2 \ .
\end{equation} 
The same differential equation can be obtained by differentiating 
(\ref{commentxor1}). With the limit condition
$s_h^*(\alpha\to\infty)=0$ we obtain back the correct expression for 
the average entropy of homogeneous systems. The entropy is equal to
$(1-\alpha)\ln 2$ at $\alpha=\alpha_c$, and becomes smaller when the
ratio decreases. This shows that the solution $b=0$ must be
preferred in this regime to the metastable $b>0$ solution.
We conclude that the cavity assumption leads to sensible results 
for homogeneous systems at all ratios
$\alpha$.

In full systems the sign $\sigma$ entering (\ref{expressu}) takes $\pm 1$
values with equal probabilities. We thus expect $p(h)$ to be an even 
distribution, and $a_+=a_-=\frac 12 (1-a_0)$. 
Remark that a solution with  $a_0<1$ cannot exist in the satifiable
phase. It would allow two added equations to impose opposite non zero biases
to the new variable {\em i.e.} to constraint this variable to take
opposite values at the same time.
Given $a_0$ we calculate from (\ref{defph}) the probability that the
field vanishes,
\begin{equation}\label{e1a}
p(0)= e^{-3\,\alpha (1-a_0)} \sum _{\ell =0} ^\infty \left[ 
\frac {3\alpha}2 (1-a_0)\right]^{2\ell} \frac 1{\ell !^2}
\end{equation}
and, in turn, derive from (\ref{conda0}) a self-consistent equation 
for $a_0$.
Numerical investigations
show that $a_0=1$ is the unique solution for $\alpha < \alpha_T=
1.167$. When $\alpha > \alpha_T$ there appears another solution with
$a_0<1$. 
The clustering and Sat/Unsat transitions are totally absent. 
This result, incompatible with the exact picture of random 3-XORSAT
exposed in Section \ref{seccluster}, shows that the simple cavity 
hypothesis does not hold for full systems. 

\subsection{Spontaneous symmetry breaking between clusters}
\label{secpin}

In the clustered phase variables are known to be strongly correlated
and the cavity assumption has to be modified. Actually from what we
have done above in the homogeneous case we guess that the independence
condition still holds if we can in some way restrict the whole space
of solutions to one cluster. To do so we explicitely
 break the symmetry between clusters as follows\cite{Pa86,Mon95}.

Let $S_i^*,i=1,\ldots, N$ be a reference solution of a full
satisfiable system $F$, and $F_h$ the corresponding homogeneous
system. We define the local gauge transform $S_i \to \hat S_i= S_i \times
S_i^*$.  $\{S\}$ is a solution of $F$ if and only if $\{\hat S\}$ is a
solution of $F_h$. As the cavity assumption is correct for the
homogeneous system we obtain the distribution of fields $\hat
h_i \ge  0$ from (\ref{defph}). Gauging back to the original spin 
configuration gives us the fields
\begin{equation}\label{gauge}
h_i = S_i^*\times \hat h_i \ .
\end{equation}
It turns out that the above fields depend only on the cluster to which
belong the reference solution.  Indeed 
for the fraction $1-b$ of the non frozen spins, $\hat h_i = 
h_i=0$. For the remaining fraction $b$ of spins in the backbone 
$\hat h_i >  \frac 12$ and $S_i^*$ has a unique value for all
solutions in the cluster (Section \ref{seccluster}). Hence
the fields $h_i$ are a function of cluster $(c)$ containing $\{S^*\}$,
and will be denoted by $h_i ^{(c)}$.

What modification has to be brought to the cavity assumption of
Section \ref{seccav1} is now clear. Given a subset $I$ of the spins
with configuration $S_I$ we define $E_F^{(c)}(S_I)$ as the GS
energy over configurations in the cluster $(c)$. Then the cavity
assumption is correct (spins in $I$ are uncorrelated) and $E_F^{(c)}$
define the fields $h_i^{(c)}$. How do we perform this restriction in
practice? A natural procedure is to break the symmetry between
clusters in an explicit manner by adding a small coupling to the
reference solution \cite{Pa86,Mo95}. 
Remark that symmetry was broken (naturally but explicitly!) in the case of
homogeneous systems when we looked for a distribution $p(h)$ with
support on positive fields only. It is a remarkable feature of XORSAT 
(rather unique among disordered systems) that symmetry between 
disordered clusters can be broken in a constructive and simple way.

The main outcome of the above discussion is that the field attached to
variable $i$ is not unique, but depends on the cluster $(c)$. We define
the distribution $p_i(h)$ of the fields attached to variable $i$ over
the clusters (with uniform weights since all clusters contain the same
number of solutions) \cite{Mon98}. The naive cavity assumption
corresponds to 
\begin{equation}
p_i(h) = \delta _{h-h_i} \ .
\end{equation}
In presence of many
clusters $p_i(h)$ is not highly concentred. From (\ref{gauge}) and the fact
that $S_i^*=\pm 1$ depending on the cluster from which we pick up the 
reference solution we find that  
\begin{equation}\label{ef}
p_i(h) = \frac 12 \left[ \delta _{h-\hat h_i} + \delta _{h+\hat h_i}\right] \ .
\end{equation}
As $\hat h_i$ is itself randomly distributed we are led to introduce
the distribution ${\cal P}$
of the field distributions $p_i(h)$. This mathematical object, ${\cal
P}(p(h))$, is the order parameter of the cavity theory in the
clustered phase \cite{Mon98,Me01y}.

\subsection{Distribution of field distributions} \label{seccavrsb}

Let us see how ${\cal P}$ can be obtained within the one-more variable
approach of Section \ref{seccav1}. A new equation  
contains two variables $S_i,S_j$ from $F$, with fields
$h_i^{(c)},h_j^{(c)}$ in each cluster $(c)$. The bias $u$ 
is a deterministic 
function of those two fields for each cluster (\ref{expressu}). We
define its distribution over clusters $\rho$. As $u$ can take three
values only and $\rho$ is an even distribution due to the randomness of 
the second member of the new equation  we may write 
\begin{equation} \label{db}
\rho (u ) = (1-\varphi_{ij}) \; \delta_u + \frac{\varphi_{ij} }2\;
\big( \delta _{u-\frac 12} +  \delta _{u+\frac 12} \big) \ .
\end{equation}
The weight $\varphi$ is a random variable which varies from pair
$(ij)$ to pair. 

What is the probability distribution {\sc p}$(\varphi)$ of $\varphi$?
Either the two variables in the pair belong to the backbone and they
are frozen in all clusters; then $u$ will be non zero and
$\varphi=1$. Or one (at least) of the two variables is not frozen and $u=0$
in all clusters, giving $\varphi=0$. We may write
\begin{equation}
\mbox{\sc p}(\varphi ) = (1-w)\; \delta _{\varphi} + w\;
\delta_{\varphi-1} \ . 
\end{equation}
From the above argument we expect $w=b^2$. Let us derive this result.

Assume we add $\ell\ge 1$ equations to our system.  For each one of
those equations a bias $u^{j}$ is drawn randomly according to distribution 
(\ref{db}). Denote by $m(\le
\ell)$ the number of those equations with parameter $\varphi=1$; $m$
is binomially distributed with probability $w$ among $\ell$.
Then $\ell-m$ biases are null, and $m$ biases are not equal to zero.
For the formula to remain satisfiable the non-zero biases must be all
positive or negative \cite{Me01y}, see  Section 
\ref{seccavapp}. Hence the distribution of the field on the new variable 
is
\begin{equation}\label{dispm}
p^m(h) = \frac 12 \left[ \delta_{h-\frac m2} + \delta _{h+\frac m2}\right] \ ,
\end{equation}
in agreement with the expected form (\ref{ef}).
The upperscript $m$ underlines that field distributions with non zero
probability are can be labelled by an integer $m$; they define
a countable set and the distribution ${\cal P}$ can be defined
as a discrete probability ${\cal P}_m$ over the set of positive
integers $m$.
The probability ${\cal P}_m$ of distribution (\ref{dispm}) 
is the convolution of 
binomial distribution for $m$ at fixed $\ell$ with the Poisson
distribution over $\ell$,
\begin{equation}\label{dispmp}
{\cal P} ^m = e^{-3\,\alpha\, w} \frac{(3\,\alpha\, w)^m
}{m!} \ .
\end{equation}
Identities (\ref{dispm},\ref{dispmp}) fully determine the distribution of field
distributions in term of a single parameter, $w$.

To close the self-consistency argument consider the two variables in
$F$ in, say, the first added equation. Call $h,h'$ their fields,
distributed according to $p_m(h), p_{m'}(h')$ for
some $m,m'$. The bias  created onto the new variable will be non
zero if $h$ and $h'$ may both take non zeros value in some clusters,
that  is, if $m$ and $m'$ are not equal to zero. 
This translates into the mathematical identity 
\begin{equation}
w = \sum _{m\ge \frac 12,m'\ge \frac 12} 
{\cal P}_m\; {\cal P}_{m'} = \big(1-{\cal
  P}_0 \big)^2 = \big(1 - e^{-3\,\alpha\, w}\big)^2
\end{equation}
from (\ref{dispmp}).
The above equation coincides with (\ref{eqb}) for $w=b^2$. Notice that
$w$ is equal to the probability $a_+$ 
that the bias is non zero in the homogeneous
case (\ref{uk}), in agreement with the discussion of Section \ref{secpin}.

It is easy to find back the expressions for the entropies of clusters,
$s_{cluster}$, and solutions in a cluster, $s_{in}$, given in Section
\ref{seccluster}.  As for the latter entropy the argument leading to
(\ref{evolsh}) can be repeated, with the modification that the second
member of the added equation is not necessarily zero but the value it
should have for the equation to be satisfied when all three variables
are frozen. Hence (\ref{evolsh}) holds with $s_{h}^*$ replaced with
$s_{in}$. As for the entropy of clusters the same argument again tells us
that, on average, half of the clusters will disappear when the 
three variables are frozen and the second member of the equation is
randomly chosen. Therefore
\begin{equation}\label{sc2}
\frac{d s_{cluster}}{d\alpha} = -b^3\, \ln 2 \ ,
\end{equation}
in agreement with equations (\ref{scluster},\ref{eqb}).
Summing differential equations (\ref{sc2}) and (\ref{evolsh}) 
for $s_{cluster}$ and $s_{in}$ respectively shows that the total
entropy of solutions is $(1-\alpha)\, \ln 2$ (Section \ref{seccluster}).


\section{Dynamical phase transitions and search algorithms}

The deep understanding of the statistical properties of
3-XORSAT makes this problem a valuable benchmark for assessing the 
performances of
various combinatorial search algorithms. At first sight, the idea
seems rather odd since 3-XORSAT is a polynomial problem. 
Interestingly most of the search procedures devised to deal with 
NP-complete problems e.g. SAT have poor performances {\em i.e.}
take exponentially long average running times
on XORSAT above some algorithmic-dependent critical ratio ... 
The purpose of this Section is to present two algorithms exhibiting
such a dynamical phase transition, and the techniques required for 
their analysis.

\subsection{Random WalkSAT (RWSAT): definition, worst-case bound}

The first algorithm we consider is the Random WalkSAT (RWSAT) 
algorithm introduced by Papadimitriou \cite{Pa92}.
RWSAT is based on the observation that a violated equation can be satisfied 
through negation of one of its variables:

{\sc
\begin{itemize}
\item   Start from a randomly chosen configuration of the 
variables. Call energy the number $E$ of unsatisfied equations.
\item While $E\ge 1$;   
\begin{itemize}
\item pick up uniformly at random one of the $E$ unsatisfied equations;
\item  pick up  uniformly at random one of its 3 variables;
\item  negate the value of this variable, update $E$;
\end{itemize}
\item Print 'Satisfiable', and Halt.
\end{itemize}
}

\vskip .3cm
Notice that, as a result of the negation, some equations 
that were satisfied may become violated. Therefore the energy is
not guaranteed to decrease with the number of steps of the algorithm.
RWSAT is able to escape from local minima
of the energy landscape, and is {\em a priori} capable of better 
performances. From the other hand, RWSAT may run forever...
A major question is how long should the algorithm be running 
before we stop thinking that the studied system has solutions
hard to find and get some confidence that there is really no 
solution. 

This question was addressed by Sch\"oning \cite{Sc00}, who
showed that RWSAT could easily be used as a one-sided
randomized algorithm \cite{Mo95}\footnote{Sch\"oning's 
original work was devoted to
the analysis of RWSAT on K-SAT, but his result holds for K-XORSAT
too.}. Consider one instance of 3-XORSAT and run RWSAT for $3N$ steps
from a randomly chosen configuration of variables. Choose again a
random initial configuration and run RWSAT another $3N$ steps, and so on
... The probability that no solution has been found after $T$ 
repetitions of this procedure though the formula is satisfiable is
\begin{equation} \label{ressch}
p_{SAT} \le \exp \left( - T \times \left(\frac 34\right)^{N +o(N)} \right) \ .
\end{equation}
Hence we obtain a probabilistic proof that the instance is not
satisfiable if the algorithm has run unsuccessfully for more than
$(\frac 43)^N$ sets of $3N$ steps. It must be clear that this result
holds for any instance, no assumption being made on the distribution
of formulas. The probability appearing in (\ref{ressch}) is on the
random choices done by RWSAT and the choices of the restart
configurations for a fixed formula.

The proof of (\ref{ressch}) can be sketched as
follows. Assume that the formula is satisfiable, and called $X^*$ one
of its solutions. Consider now the (extensive)  Hamming
distance between the solution and the configuration $X$ of variables 
produced by RWSAT at some instant. After each step only one variable is
changed so $D$ changes into $D+1$ (bad move) or $D-1$ (good move).
Call $x,y,z$ the variables in the equation which was not satisfied by
$X$. One or three of those variables have opposite values in $X^*$.
In the latter case the flip is always a good move;
in the former case the good move happens with probability  $\frac 13$ and
a bad move with probability $\frac 23$. On the overall the probability of
a good move is $\frac 13$ at least. 

Think of $D$ has the position of a random walker on the $[0;N]$ 
segment. Initially the position of the walker is a binomial variable, 
centered in $\frac N2$. At each step the walker moves to the left 
with probability $\frac 13$, and to the right with probability $\frac 23$. 
We look for the probability $\rho$ that the walker is absorbed by the 
boundary $D=0$ after  $S$ steps. A standard calculation shows that
$\rho$ is maximal for $S=3N$, with the value $\rho\simeq
(\frac 34)^N$. After $T$ repetitions the probability of not having been
absorbed is $(1-\rho)^T<\exp(-\rho\,T)$, hence (\ref{ressch}).
The proof can be easily extended to $K$-XORSAT with higher values of $K$.
The number of repetitions necessary to prove unsatisifiability
scales as $(\frac{2(K-1)}K)^N$; it is essentially equal to $2^N$ for large 
$K$, showing that RWSAT does not beat exhaustive search in this limit. 

\subsection{Dynamical transition of RWSAT on random XORSAT instances}
\label{secrwsat}

\begin{figure}
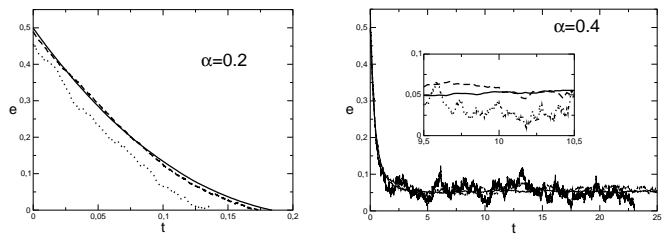

\begin{center}
\includegraphics[width=110pt,angle=0]{fig521.eps}
\hskip.5cm
\includegraphics[width=120pt,angle=0]{fig522.eps}
\end{center}
\caption{Fraction $e$ of unsatisfied equations as a function of time $t$
(number of steps divided by $N$) during the operation of RWSAT on a
random 3-XORSAT formula at ratio $\alpha =0.2$ (left)
and $\alpha =0.4$ (right) with $N=10^3$ (dotted), $10^4$ (dashed), $10^5$
(full curve) variables.   Note the difference
of horizontal scales between the two figures. Inset:  blow
up of the $t\in[9.5;10.5]$ region; the amplitude of fluctuations around 
the plateau decreases with increasing size $N$. }
\label{wsat_phen}
\end{figure}

\begin{figure}
\begin{center}
A\includegraphics[width=110pt,angle=-90]{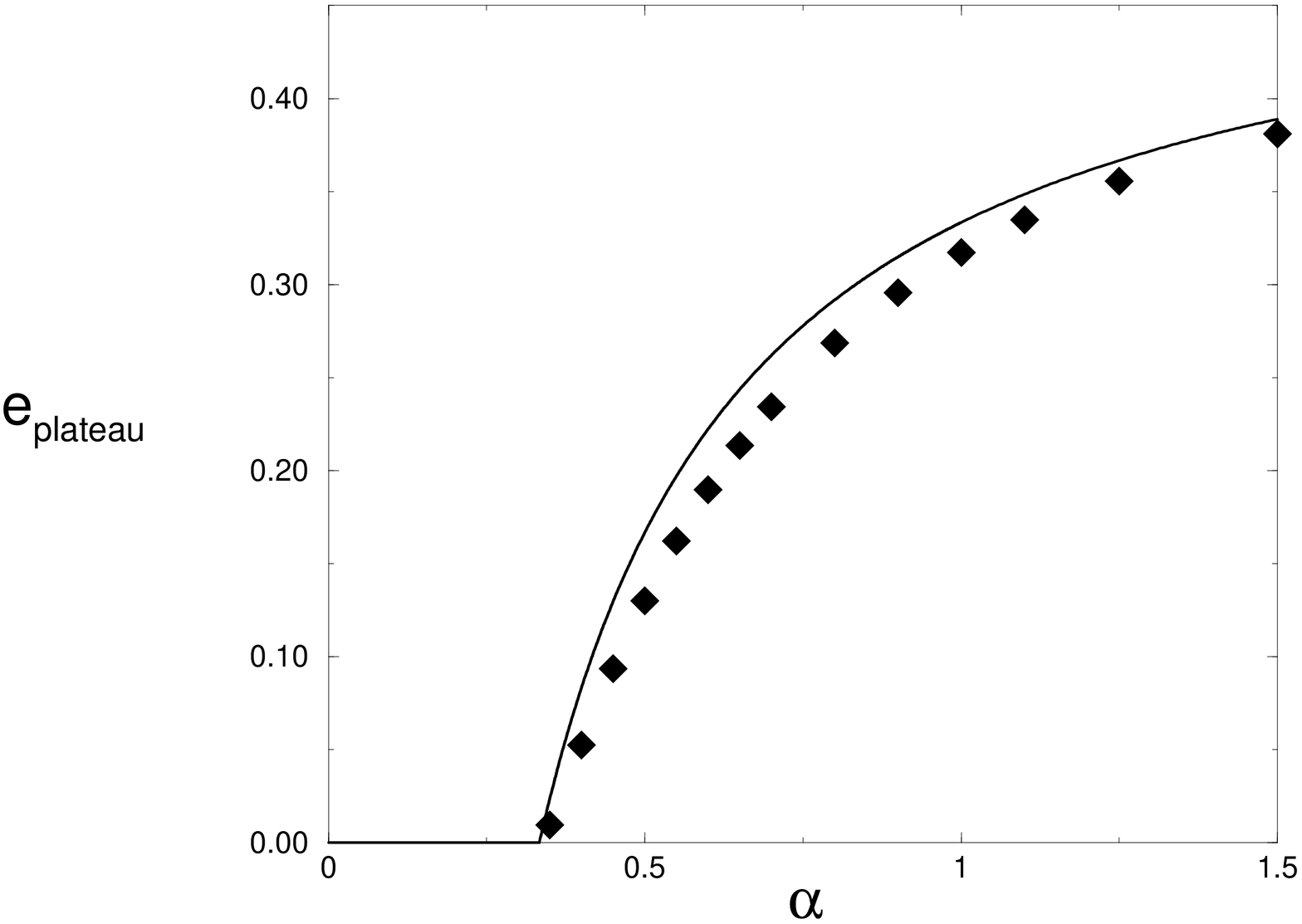}
B\includegraphics[width=110pt,angle=-90]{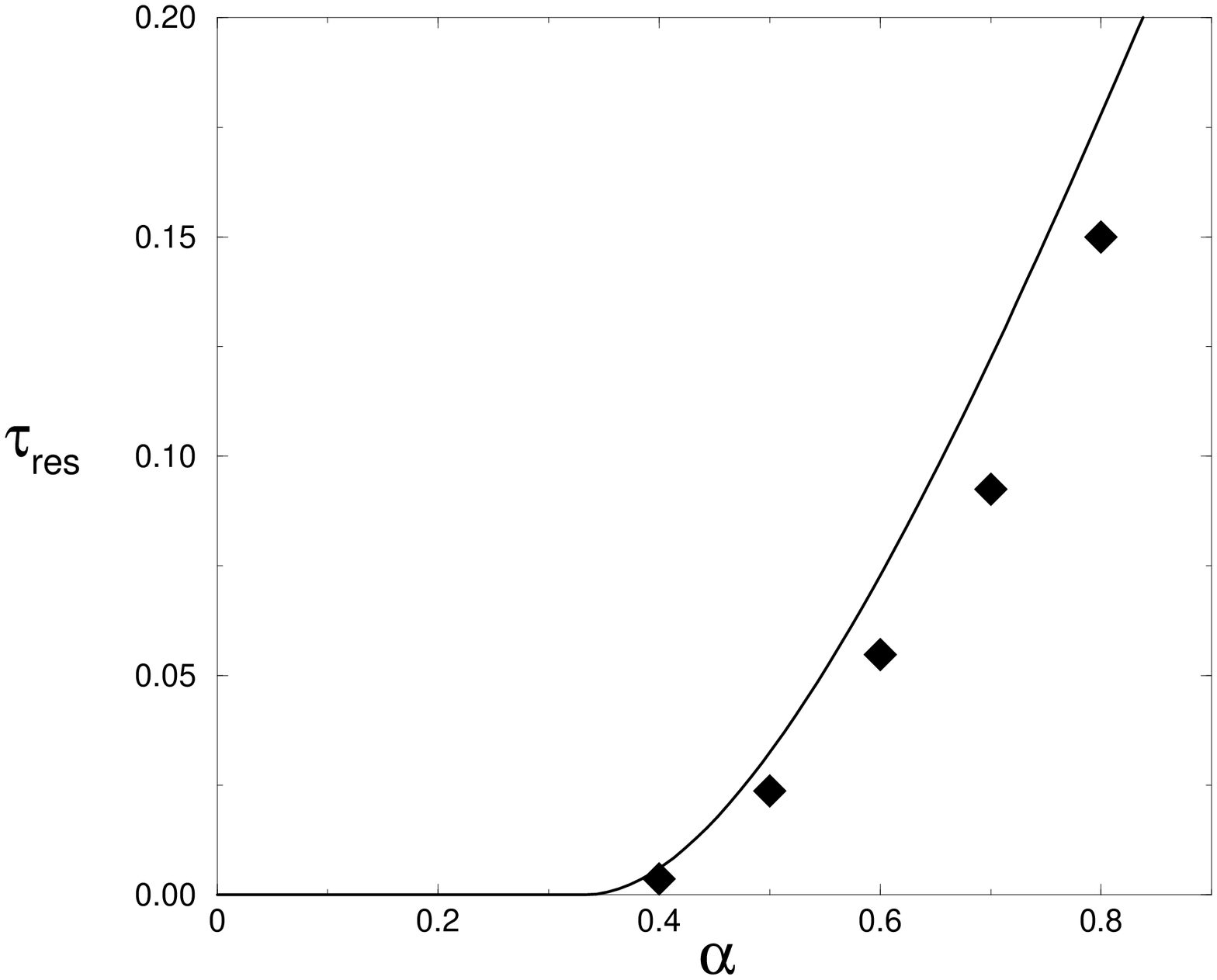}
\end{center}
\caption{Fraction $e _{plateau}$ of unsatisfied equations on 
the plateau  ({\bf A}) and logarithm
$\tau _{res}$ of the average resolution time divided by $N$ ({\bf B})  as a 
function of the ratio $\alpha$ of equations per variable. 
Diamonds are the output of numerical experiments, and have been obtained 
through average of data from simulations over 1,000 systems and runs of RWSAT
for various sizes $N$, and  extrapolation to $N\to\infty$ \cite{Se03}.
Full lines are theoretical approximations (\ref{eplateau}),(\ref{trestheo}).} 
\label{wsat_plateau}
\end{figure}

Result (\ref{ressch}) is true for any instance; what is the typical
situation for random systems? Numerical experiments 
indicate that there is critical value of the ratio of
equations per variables, $\alpha_E\simeq 0.33$, hereafter referred to 
as dynamical  threshold, separating two regimes:
\begin{itemize}
\item for $\alpha < \alpha _E$, RWSAT generally finds a solution very
quickly, namely with a number of flips growing linearly with the
number of variables $N$\footnote{A proof of this statement was
  obtained by \cite{Al02} for the random SAT model.}.  
Figure~\ref{wsat_phen} shows the plot of
the fraction $e$ of unsatisfied clauses as a function of the time
(number of steps) $T$ for one randomly drawn system with ratio
$\alpha=0.2$ and $N=500$ variables. The curve shows a fast decrease from
the initial value ($e(T=0)=\frac 12$ 
independently of $\alpha$ for large values of $N$, but deviations can be
found at small sizes, see Figure~\ref{wsat_phen}) down to zero on a time scale
of the order of $N$\footnote{This decrease characterises the overall operation
of RWSAT. A precise look at the $e(T)$ curve reveals that the energy $e$
may occasionally increase.}. The resolution time $T_{res}$ depends
both on the system of equations under consideration and the choices of
the algorithm; its average value scales as 
\begin{equation}
\langle T_{res} \rangle  =  N\, t _{res} + o(N) \ .
\end{equation}
where $t_{res}$ is an increasing function of 
$\alpha$\footnote{On intuitive grounds, as a step of the algorithm can satisfy 
$\theta(1)$ equations at a time, we expect the average value of 
$T_{res}$ to be of the order of the number $M$ of equations at least. 
Thus $t_{res}$ should grow at least linearly with $\alpha$. Experiments
shows that the growth is in fact more than linear.}.

\item for systems with ratios of equations per variable in the 
$\alpha_E<\alpha <\alpha_c$ range, the initial relaxation regime taking place 
on the $O(N)$ time scale does not allow RWSAT
to reach a solution (Figure~\ref{wsat_phen}B). The fraction
$e$ of unsat equations then fluctuates around some plateau value
$e_{plateau}$ 
for a very long time. Fluctuations are smaller and smaller (and the height of
the plateau better and better defined) as the size $N$ increases.
As a result of fluctuations, the fraction $e$ of unsatisfied equations
may temporarily either increase or decrease. When a fluctuation happens to 
drive RWSAT to $e=0$, a solution is found and the algorithm stops.
The corresponding resolution time, $T_{res}$, is stochastic;
numerical experiments for different sizes $N$ indicate that its
expectation value scale as
\begin{equation} \label{deftaures}
\langle T_{res} \rangle= \exp( N\, \tau _{res}+o(N)) \ .
\end{equation}
where the coefficient $\tau_{res}$ is an increasing function of $\alpha$.
The plateau energy $e_{plateau}$  and the logarithm $\tau _{res}$ of
the resolution time are shown in Figure~\ref{wsat_plateau}.  
\end{itemize}

Notice that the dynamical threshold $\alpha_E$ 
above which the plateau energy is positive is strictly smaller than the
critical threshold $\alpha_c\simeq 0.918$, where systems go from 
satisfiable with
high probability to unsatisfiable with high probability.  In the
intermediate range $\alpha_E<\alpha<\alpha_c$, systems are almost
surely satisfiable but RWSAT needs an exponentially large time to
prove so. The reason is that  RWSAT remains trapped at a high energy level 
(plateau of Figure~\ref{wsat_plateau}) for an exponentially large
time. The emergence of metastability can be qualitatively studied with simple 
tools we now expose. 

\subsection{Approximate theory for the metastable plateau and the escape time}
\label{secescape}

Assume that after $T$ steps of the algorithm the energy (number of unsatisfied
equations) is $E_T\ge 1$. Then pick up an unsatisfied equation, say, 
$C$, and a variable in $C$, say, $x$, and flip it. The energy after the flip is
\begin{equation} \label{evole}
E_{T+1} = E_T -1 -U+S \ ,
\end{equation}
where $S$ (respectively $U$) is the number of equations including $x$ 
which were satisfied (resp. unsatisfied after exclusion of equation $C$) 
prior to the flip. $S$ and $U$ are random variables with binomial 
distributions,
\begin{eqnarray}
\mbox{Proba}[U] &=& {E_T-1 \choose U} \left( \frac 3N\right)^U
\left( 1-\frac 3N\right)^{E_T-1-U} \ , \nonumber \\
\mbox{Proba}[S] &=& {M-E_T \choose S} \left( \frac 3N\right)^S
\left( 1-\frac 3N\right)^{M-E_T-S} \ . 
\end{eqnarray}
where the probabilities are intended over the formula content.
Taking the average evolution equation (\ref{evole}) we obtain
\begin{equation} \label{evole2}
\langle E_{T+1}\rangle =\langle E_T\rangle -1 -\frac 3N \big(
 \langle E_T\rangle -1\big) +\frac 3N \big( M - \langle E_T\rangle\big) \ .
\end{equation}
The above equation is exact. It is now tempting to iterate it with time, from
the initial condition $\langle E_{T=0}\rangle = \frac M2$. This is what we do
hereafter but one should realize that this procedure is not
correct from a mathematical standpoint. The catch is that one is allowed to
average over the formula only once, and certainly not at each time step 
of the algorithm. Evolution
equation (\ref{evole2}) amounts to redraw randomly the instance
at each time step, conditioned to the energy. This approximation
nevertheless allows
us to write down a simple equation for $\langle E_T\rangle$, which
captures much of the true behaviour of RWSAT.

The next step in our analysis is the large size, large time limit. As
the energy can typically change by a quantity of the order of unity 
in one time step we expect the fraction of unsatisfied equations to
vary of a time scale of the order of $N$,
\begin{equation}\label{evola}
\langle E_T\rangle = M\; e\left( \frac TM =t\right) \ ,
\end{equation}
for some smooth function $e(t)$ of the reduced time $t$.
Finite difference equation (\ref{evole2}) turns into a differential equation 
after insertion of (\ref{evola}),
\begin{equation}\label{evole3}
\frac{de}{dt} = - 1 + 3\alpha \, ( 1 -2\, e) \ ,
\end{equation}
with the initial condition $e(0)=\frac 12$. Clearly (\ref{evole3}) makes
sense as long as $e>0$; if $e$ vanishes the algorithm stops.
Resolution of (\ref{evole3}) shows the following scenario. 
If $\alpha$ is smaller than
\begin{equation}
\alpha _E = \frac 13 \ ,
\end{equation}
the fraction $e$ of unsatisfied equations quickly decreases, and
vanishes at some time $t_{res}(\alpha)$. This regime corresponds to
a successfull action of RWSAT in a $O(N)$ number of steps.
$t_{res}$ is an increasing function of $\alpha$ which diverges as
$\alpha \to\alpha_E$. Above this critical ratio $e$ shows a different
behaviour: after a decreasing transient regime $e$ saturates to a
positive plateau value 
\begin{equation}\label{eplateau}
e_{plateau} (\alpha) = \frac 12 \left( 1 - \frac {\alpha_E}{\alpha}\right) 
\ .
\end{equation}
The value of the plateau energy is compared to numerics in Figure
\ref{wsat_plateau}A. The agreement on the location of the dynamical 
threshold $\alpha_E$ as well as the plateau energy are satisfactory.

The remaining point is to understand how RWSAT finally finds a solution 
when $\alpha >\alpha_E$. The above theory, based on taking the 
$N\to\infty$ limit first, washes out the fluctuations of the energy around
its metastable value, of crucial importance for resolution
\cite{Se03}. To take into account these fluctuations let us define
the probability $Q_{plateau}(E)$ that the energy takes value $E$ 
in the plateau regime of Figure \ref{wsat_phen}B. A stationary 
distribution is well defined if we discard the initial transient
regime (choose large $t$) and collect values for $E$ on exponentially 
large--in--$N$ time scales. The procedure is standard in the
study of long-time metastable states.

Within our draw-instance-at-each--step approximation we may write
a self-consistent equation for the stationary distribution
of energies,  
\begin{equation} \label{statq}
Q_{plateau}(E) = \sum _{U,S} \mbox{Proba}[U]\, \mbox{Proba}[S]\,
\, Q_{plateau}(E+1+U-S) 
\end{equation}
where the meaning of $U,S$ was explained right
after (\ref{evole}). From Section \ref{secrwsat} we expect fluctuations
to decreases sharply with the system size. A reasonable guess for the
scaling of the distribution with $M$ is 
\begin{equation}
Q_{plateau}(E) = 
\exp\left[ -M\; \omega \left(\frac EM =e\right) +o(M)\right]
\end{equation}
where $\omega$ is the rate function associated to the fraction
of unsatisfied equations. Plugging the above Ansatz into (\ref{statq})
and taking the large $M$ limit we find that $\omega$ fulfills the
following differential equation 
\begin{equation}\label{pde1}
F\left( \frac{\partial \omega}{\partial e},e\right) =0\ ,
\end{equation}
where $F(x,y)=3\alpha y(e^{-x}-1) +3\alpha (1-y) (e^x-1) -x$.
This equation has to be solved with the condition 
$\omega(e_{plateau}) =0$. 

An analytical solution can be found for (\ref{pde1}) when we restrict
to the vicinity of the dynamical transition {\em i.e.} to small
values of $\omega$. 
Expanding $F$ to the second order in its first argument and solving
(\ref{pde1}) we obtain
\begin{equation} \label{ome}
\omega(e) \simeq 2 \, (e - e_{plateau})^2\ , 
\end{equation}
where $e_{plateau}$ is defined in (\ref{eplateau}).

What happens when time increases is now clear. Assume we have run RWSAT
up to time $t\sim e^{M\tau}$. Then configurations with energy $e$ such that
$\omega (e) <\tau$ have been visited many times and are 'equilibrated'
with probability (\ref{statq}), (\ref{ome}). Configurations with
energies outside the band $e_{plateau} \pm \sqrt{\tau /2}$ are not
accessible. When the time scales reaches
\begin{equation} \label{trestheo}
\tau _{res} = \omega(0) \simeq \frac 12 \left( 1 - \frac 1{3\alpha}\right)^2  
\ ,
\end{equation}
zero energy configurations are encountered, and RWSAT comes to a stop.
The agreement between the theoretical estimate (\ref{trestheo}) and
the numerical findings (\ref{deftaures}) visible in Figure \ref{wsat_plateau}B
is acceptable in regard to the crudeness of the approximation done.

\subsection{Davis-Putnam-Loveland-Logemann (DPLL) algorithm}\label{secdpll}

The second procedure is the Davis-Putnam-Loveland-Logemann
(DPLL) algorithm \cite{dpll}. 
Contrary to RWSAT DPLL can provide exact proofs for
unsatisfiability. The procedure, widely used in practice, is based on 
the trial-and-error principle. Variables are assigned
according to some heuristic rule (split step), and
equations involving those variables simplified. If  an
equation involving a single variable (unit-equation)  appears
its variable is chosen accordingly prior to any other 
heuristic assignment
(unit-propagation).  If a contradiction is found (two opposite
unit-equations) DPLL backtracks to the last heuristically assigned 
variable, flips it, and resumes the search process. 
The procedure halts either when all equations have been
satisfied (a solution is then found), 
or when all possible values for the variables have been tried in vane and
found to be contradictory (a proof of unsatisfiability
is then obtained). 

DPLL can be described as a recursive 
function of the variable assignment $A$. 
Given a system $S$  DPLL is first called
with the empty assignment $A=\emptyset$:
\vskip .3cm
{\sc Procedure DPLL[$A$]
\begin{itemize}
\item Let $S_A$ be what is left from $S$ given variable assignment $A$;
\item if $S_A$ is empty,  Print `Satisfiable'; Halt;
\item If $S_A$ contains a violated equation, Print `Contradiction', Return; 
{\em (backtracking)}  
\item Otherwise, let $U$ be the set of unit-equations in $S_A$;     
\begin{itemize}
\item If $U \ne \emptyset$, pick-up one of the equations in
$U$, say, $e$, and call DPLL[A$\cup \{e\}$];
{\em (unit-propagation)}
\item if $U=\emptyset$, choose a not-yet-assigned
variable, say, $x$, and its value $v$ according to some heuristic rule, and
 call DPLL[A$\cup \{x=v\}$], then
DPLL[A$\cup \{x=\bar v\}$]; {\em (variable splitting)}
\end{itemize}
\end{itemize}
}

\vskip .3cm
Rules for assigning variables in the absence of unit-equations are heuristic
in that they aim at doing good assumptions {\em i.e.} diminishing as much 
as possible the search process to come from limited information
about the current system of equations. Of course, perfect heuristic 
do exist:  trying all possible values for not-yet-assigned variables
would ensure that no wrong guess is ever done!  But the time required 
would be exponentially long. In practice, heuristics have to make 
their decision in polynomial time. Two simple splitting heuristics 
are:
\begin{itemize}
\item[$\diamond$] \underline{UC:} 
choose at random and uniformly any unset variable,
and assign it to 0 or 1 with equal probabilities ($\frac 12$). 
\item[$\diamond$]  \underline{GUC:}
choose at random and uniformly any equation with minimal length {\em i.e.}
involving 2 variables if any, or 3 variables otherwise. Pick up 
at random and uniformly  one its variable, and assign it to 0 or 1 
with equal probabilities ($\frac 12$). 
\end{itemize}  
UC, which stands for unit-clause \cite{Ch90}, 
amounts to make a random guess and is the simplest possible heuristic.
GUC (Generalized UC) 
is more clever: each time a split is done from an equation with 2 
variables, this equation is turned into a unit-equation, and
eliminated through unit-propagation. 
In the following, we call DPLL-UC and DPLL-GUC the variants of
DPLL based on the UC and GUC heuristics respectively.

\begin{figure}
\begin{center}
\includegraphics[width=110pt,angle=-90]{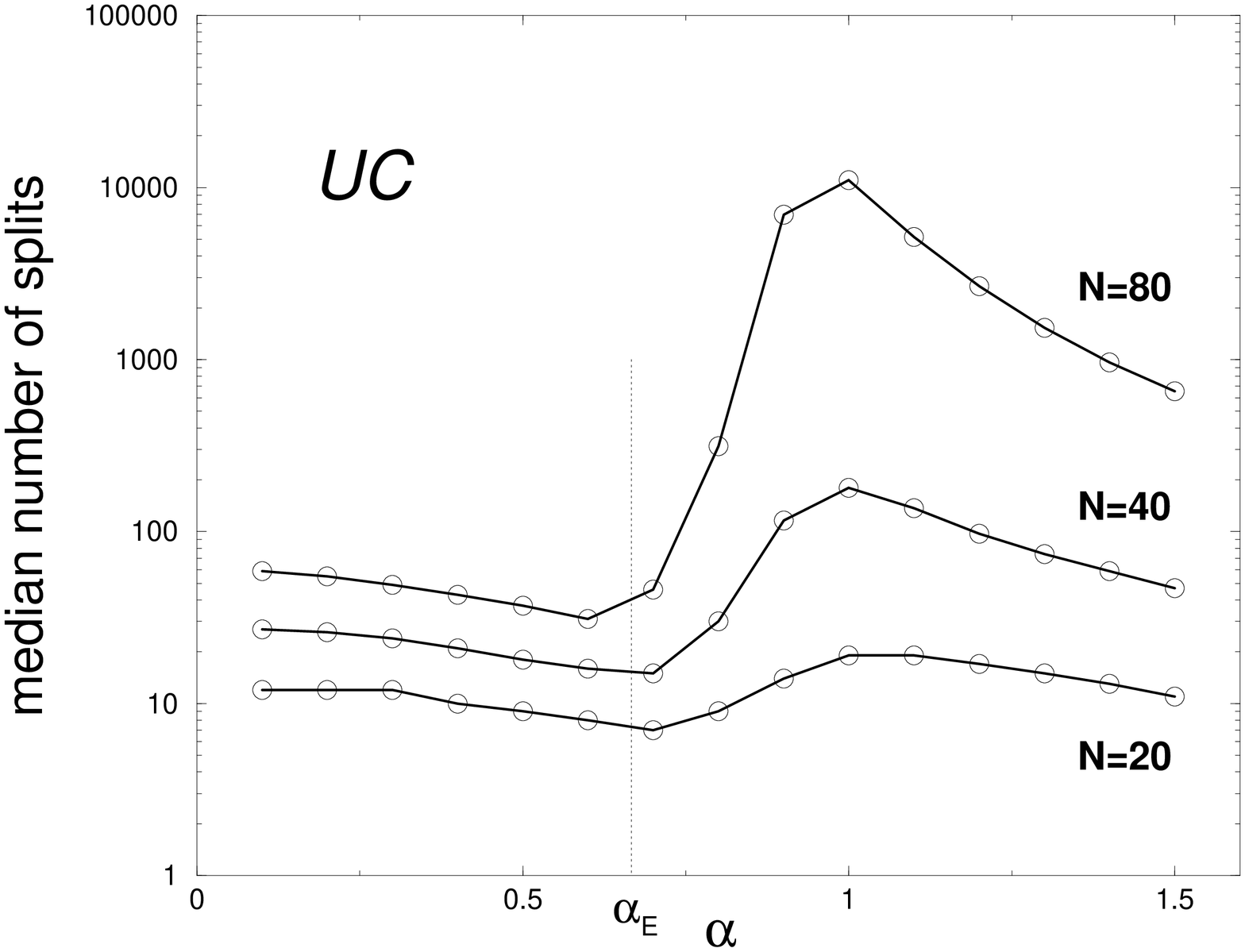}
\hskip .3cm 
\includegraphics[width=110pt,angle=-90]{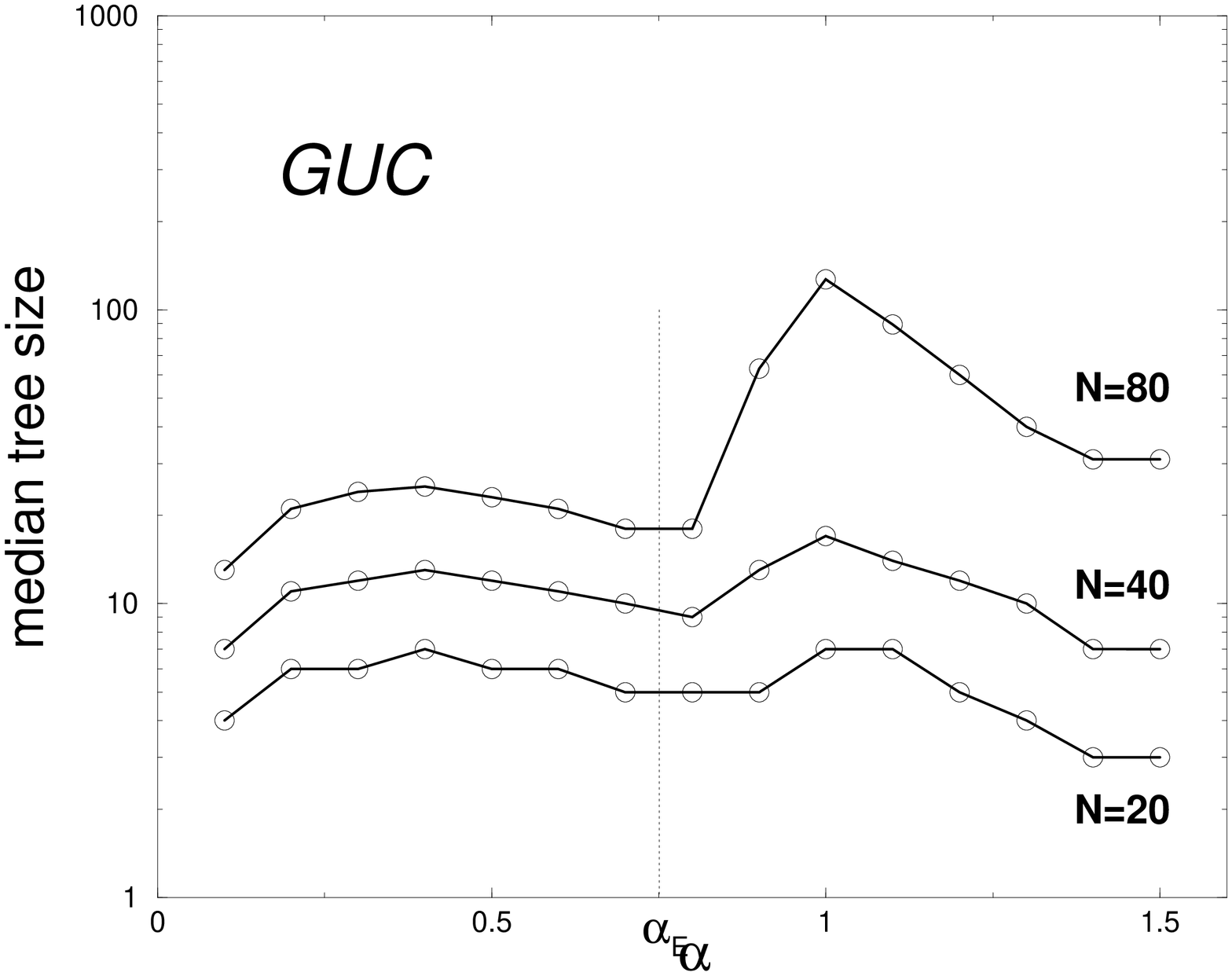}
\end{center}
\caption{Median number of splits required by DPLL 
with the UC (left) and GUC (right) heuristics as a function of the ratio 
$\alpha$, and for $N=20,40,60$ variables (from bottom to
top). Data have been extracted from the resolution
of 10,000 randomly drawn systems; continuous lines are guidelines for the 
eye. Note the difference of (logarithmic) scale between UC and GUC
curves showing that DPLL-GUC is much more efficient than DPLL-UC. 
The polynomial/exponential transition is located at ratios
$\alpha_E=\frac 23$ and $\alpha_E=0.7507...$ for UC and GUC respectively.}
\label{dpllxorsat}
\end{figure}

A  measure of the computational effort required
by DPLL is the number $T_{split}$ 
of variable splittings. This number varies from
system to system (at fixed number $N$ of variables and ratio $\alpha$), and
from run to run of DPLL due to the stochasticity introduced by
the heuristic rule. The outcome of numerical experiments for
the  median number of splits\footnote{The median is more
representative of the typical value of the number of splits than the
expectation value, since the latter may be dominated by huge and
unlikely samples, see discussion of Section \ref{secselfav}.}.
For a given size $N$ $T_{split}$ shows a maximum located
around $\alpha\simeq \alpha_c$. If one fixes $\alpha$ 
$T_{split}$ is an increasing function of the size $N$; 
numerical data 
support the existence of a dynamical threshold, $\alpha_E$, separating
linear and exponential scalings in $N$,
\begin{equation} \label{mesure}
T_{split}\sim \left\{\begin{array} {c c c}
 N\, t _{split} + o(N) &\hbox{\rm if} & \alpha < \alpha_E \\ 
\exp(  N\, \tau _{split}+o(N) ) &\hbox{\rm if} & \alpha > \alpha_E 
\end{array} \right. \ ,
\end{equation}
where $t_{split}$ and $\tau _{split}$ are functions of the ratio $\alpha$.
The value of the dynamical threshold can be derived from theoretical
calculations shown in Section \ref{secuc} and is equal to $\alpha_E=\frac 23$ 
and $\alpha_E\simeq 0.7507...$ for UC and GUC heuristics respectively. 
Three dynamical regimes are therefored identified \cite{Co01,Ac02}:
\begin{itemize}
\item \underline{Linear \& satisfiable phase ($\alpha<\alpha_E$):} systems with
small ratios are solved with essentially no backtracking. A solution
is found after $O(N)$ splits. 
\item \underline{Exponential\& satisfiable phase ($\alpha_E<\alpha
<\alpha_c$):}
systems with ratios slightly below threshold have solutions, but 
DPLL generally requires an exponential number of splits to find
one of them. An explanation for this drastic breakdown of performances 
will be given in Section \ref{secuc}.
\item \underline{Exponential \& unsatisfiable phase ($\alpha>
\alpha_c$):} finally,
finding a proof of unsatisfiability typically requires an exponentially
large number of splits \cite{Ch88}. Note that, as $\alpha$ gets higher 
and higher, each variable assignment affects more and more equations 
(of the order of $\alpha$), and contradictions
are detected earlier and earlier. Rigorous calculations show that
$\tau_{split} \sim \frac 1\alpha$\cite{beame}, 
and the computational effort decreases
with increasing $\alpha$ (Figure~\ref{dpllxorsat}). The median number 
of splits 
is considerably smaller for DPLL-GUC than for DPLL-UC, a result expected 
from the advantages of GUC against UC discussed above.
\end{itemize}
 
\subsection{Linear phase: resolution trajectories in the $2+p$-XORSAT
phase diagram}\label{secuc}

\begin{figure}
\begin{center}
\includegraphics[width=160pt,angle=0]{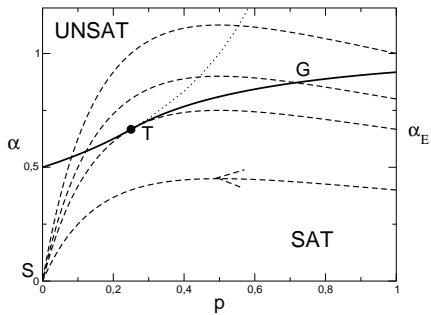}
\caption{Phase diagram of $2+p$-XORSAT and dynamical trajectories of DPLL.
The threshold line $\alpha_c (p)$ (bold full line) separates sat
from unsat  phases. Departure points for DPLL trajectories are located on the
3-XORSAT vertical axis with ratios $.4,\frac 23,.8,1.$ from
bottom to top. The arrow indicates the direction of motion along trajectories
parametrized by the fraction $t$ of variables set by DPLL.  For small
ratios $\alpha < \alpha _E$ ($=\frac 23$ for the UC heuristic)
trajectories remain
confined in the sat phase, end in S of coordinates $(0,0)$, where a
solution is found. At $\alpha_E$ the trajectory hits
tangentially the threshold line in T of coordinates $(\frac 14,\frac 23)$. 
When $\alpha >\alpha _E$ the
trajectories intersect the threshold line at some point G (which depends
on $\alpha$), and stops before hitting the
$\alpha _D (p)$ dotted line (\ref{defcon}). 
After massive backtracking DPLL will find a solution;  G corresponds
to the highest node in the search tree.}
\label{diag2+p}
\end{center}
\end{figure}

Action of DPLL on an instance of 3-XORSAT causes changes to the
numbers of variables and equationses, and thus to the ratio
$\alpha$.  Furthermore DPLL turns equations with 3 variables into
equation with 2 variables. A mixed $2+p$-XORSAT distribution, where $p$ 
is the fraction of 3-equations and $\alpha$ the ratio of the
total number of 2- and 3- equations over the number of variables
can be used to model what remains of the input 
system\footnote{Equations with a single variable
are created too, but are eliminated through unit-propagation. When 
a heuristic assignment has to be made the system is a mixture of
equations with 2 and 3 variables only.}. Repeating the calculations of
Section \ref{secreplicas} for the $2+p$-XORSAT models we derive the
phase diagram of Figure \ref{diag2+p}. The Sat/Unsat critical
line $\alpha_c(p)$ separates the satisfiable from the unsatisfiable 
phases. For $p \le p_0 = \frac 14$ {\em i.e.} to the left of
point T, the threshold line coincides with the percolation 
transition as in the 2-XORSAT model, and is given by 
$\alpha _c(p)=\frac 1{2(1-p)}$. 
For $p>p_0$ an intermediate clustered phase is found as in the
3-XORSAT model, and the threshold coincides with the vanishing of the
cluster entropy $s_{cluster}$ (Section \ref{seccluster}).

The phase diagram of 2+p-XORSAT is the natural space in which DPLL
dynamic takes place. An input 3-XORSAT instance with ratio $\alpha$ shows
up on the right vertical boundary of Figure~\ref{diag2+p} as a point of
coordinates $(p=1,\alpha )$. Under the action of DPLL the
representative point moves aside from the 3-XORSAT axis and follows a
trajectory, very much alike real-space renormalization, which 
depends on the splitting heuristic. Trajectories enjoy two essential
features \cite{Ac01}. First the representative point
of the system treated by DPLL does not `leave' the 2+p-XORSAT phase
diagram. In other words, the instance is, at any stage of the search 
process, uniformly distributed from the 2+p-XORSAT 
distribution conditioned to its equation per variable ratio $\alpha$ and 
fraction $p$ of 3-equations. This assumption is not true for
all heuristics of split, but holds for UC and 
GUC\cite{Ch90}\footnote{Analysis of more sophisticated heuristics 
e.g. based on the number of occurences of variables require to handle 
more complex instance distributions \cite{Ka02}.}.
Secondly, the trajectory followed by an instance in the course of
resolution is a stochastic object, due to the randomness of 
the instance and of the assignments done by DPLL. 
In the large size limit ($N\to\infty$) the trajectory becomes 
self-averageing {\em i.e.}
concentrated around its average locus in the 2+p-XORSAT phase 
diagram \cite{Wo95}. We will come back below on this concentration 
phenomenon.

\begin{figure}
\begin{center}
\includegraphics[width=160pt,angle=0]{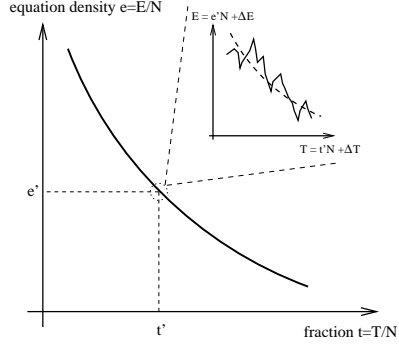}
\caption{Deterministic versus stochastic dynamics of the equation
population $E$ as a function of the number of steps $T$ of the algorithm. 
On the slow time scale (fraction $t=T/N$) the density
$e=E/N$ of (2- or 3-) equations varies smoothly according to a deterministic 
law. Blowing up of the dynamics around some point $t',e'$ 
shows the existence of small and fast fluctuations around this trajectory.
Fluctuations are stochastic but their probability distribution 
depends upon the slow variables $t',e'$ only.}
\label{deterstoch}
\end{center}
\end{figure}

Let $\alpha_0$ denote the equation per variable 
ratio of the 3-XORSAT instance to be solved. 
We call $E_j (T)$ the number of $j$--equations (including $j$ variables)
after $T$ variables have been assigned by the solving procedure. 
$T$ will be called hereafter `time', not to be confused with the 
computational effort. At time $T=0$ we
have $E_3 (0)= \alpha_0 N$, $E_2 (0)=E_1 (0)=0$. 
Assume that the variable $x$ assigned at time $T$ is chosen through
unit-propagation, that is, independently of the $j$-equation content.
Call $n_j(T)$ the number of occurrences of $x$ in 
$j$-equations ($j=2,3$). The evolution equations for 
the populations of 2-,3-equations read
\begin{equation}
E_3 (T+1) = E_3 (T) - n_3(T)\ , \quad E_2(T+1)=E_2(T) - n_2(T)+ n_3(T)   \ .
\label{evolsto}
\end{equation}
Flows $n_2,n_3$ are of course random variables that depend
on the instance under consideration at time $T$, and on the choice of
variable  done by DPLL. What are their distributions? At time $T$
there remain $N-T$ untouched variables; $x$ appears in any of
the $E_j(T)$ $j$-equation with probability $p_j = \frac j{N-T}$, 
independently of the other equations. In the
large $N$ limit and at fixed fraction of assigned variables, 
$t=\frac TN$, the binomial distribution
converges to a Poisson law with mean
\begin{equation} \label{avpois}
\langle n_j \rangle _T=\frac{j\, e_j}{1-t} \qquad \mbox{where} \qquad
e_j= \frac{E_j(T)}N
\end{equation}
is the density of $j$-equations at time $T$. The key remark is that,
when $N\to\infty$, $e_j$ is a slowly varying and non stochastic 
quantity and is a function of the fraction 
$t=\frac TN$ rather than $T$ itself. Let us iterate (\ref{evolsto})
between times $T_0=t\,N$ and $T_0+ \Delta T$ where $1\ll \Delta T
\ll N$ e.g. $\Delta T =O(\sqrt N)$. Then the change $\Delta E_3$ in the number 
of $3$-equations is (minus) the sum of the stochastic variables
$n_j(T)$ for $T=T_0,T_0+1, \ldots,T_0+\Delta T$. As these variables
are uncorrelated Poisson variables with $O(1)$ mean (\ref{avpois}) 
$\Delta E_3$ will be of the order of $\Delta T$, and the change in
the density $e_3$ will be of order of $\Delta T/N \to 0$. Applying
central limit theorem $\Delta E_3/\Delta T$ will be almost surely
equal to $-\langle n_3\rangle_t$ 
given by (\ref{avpois}) and with the
equation density measured at reduced time $t$. The argument can be
extended to 2-equations, and we conclude that
$e_2,e_3$ are deterministic (self-averaging) quantities obeying
the two coupled differential equations\cite{Ch90}
\begin{equation}
\frac{de_3}{dt}(t) = -\frac{3\, e_3}{1-t} \quad
, \qquad \frac{de_3}{dt}(t) = \frac{3\,
  e_3}{1-t}  - \frac{2\, e_2}{1-t} \ .
\end{equation}
Those equations, together with the initial condition $e_3(0)=\alpha
_0$, $e_2(0)=0$ can be easily solved,
\begin{equation}\label{sole23}
e_3(t) = \alpha_0 (1-t)^3 \quad , \qquad e_2(t)= 3\,\alpha_0
\,t\,(1-t)^2 \ .
\end{equation}
To sum up, the dynamical evolution of the equation populations may be 
seen as a slow  and deterministic evolution of the 
equation densities to which are superimposed fast, small fluctuations. 
The distribution of the fluctuations adiabatically follows 
the slow trajectory. This scenario is pictured in Figure~\ref{deterstoch}.

Expressions (\ref{sole23}) for the equation densities allow us to draw
the resolution trajectories corresponding to the action of DPLL on a
3-XORSAT instance. Initially the instance is represented by a point
with coordinates $(p=1,\alpha=\alpha_0)$ in Figure \ref{diag2+p}. As
more and more variables are assigned the representative point moves
away from the rightmost vertical axis. After a fraction $t$ of
variables have been assigned the coordinates of the point are
\begin{equation}
p(t) = \frac{e_3}{e_2+e_3}=\frac{1-t}{1+2t}\ , \quad 
\alpha(t) = \frac{e_2+e_3}{1-t}=\alpha_0(1-t)(1+2t) \ .
\end{equation}
Trajectories corresponding to various initial ratios are shown in
Figure \ref{diag2+p}.
For small ratios $\alpha_0 < \alpha _E$ trajectories remain
confined in the sat phase, end in S of coordinates $(0,0)$, where a
solution is found. At $\alpha_E$ ($=\frac 23$ for the UC heuristic), 
the single branch trajectory hits
tangentially the threshold line in T of coordinates $(\frac 14,\frac 23)$. 
When $\alpha_0 >  \alpha _E$ the trajectories enter the Unsat phase,
meaning that DPLL has turned a satisfiable instance (if $\alpha _0 <
\alpha _c$) into an unsatisfiable one as a result of poor assignments.
It is natural to expect that $\alpha_E$ is the highest ratio at which 
DPLL succeeds in finding a solution without resorting to much backtracking.

\subsection{Dynamics of unit-equations and universality}\label{secuni}

The trajectories we have derived in the previous Section are correct 
provided no contradiction emerges. But
contradictions may happen as soon as there are $E_1=2$
unit-equations, and are all the more likely than $E_1$ is large. 
Actually the set of 1-equations form a 1-XORSAT instance which is
unsatisfiable with a finite probability as soon as $E_1$ is of the order
of $\sqrt N$ from the results of Section \ref{secsf1}.
Assume now that $E_1 (T)\ll N$ after $T$ variables have been assigned, 
what is the probability $\rho_T$ that no contradiction
emerges when the $T^{th}$ variable is assigned by DPLL? This probability is
clearly one when $E_1=0$. When $E_1\ge 1$ we pick up a 1-equation,
say, $x_{6}=1$, and wonder whether the
opposite 1-equation, $x_6=0$, is present among the $(E_1-1)$ 1-equations
 left. As equations are uniformly distributed over the set
of $N-T$ untouched variables 
\begin{equation}\label{probanocttot}
\rho_T = \left( 1 - \frac 1{2 (N-T)} \right) ^{\max( E_1(T) -1,0)}  
\ .
\end{equation}
The presence of the $\max$ in the above equation ensures it remains
correct even in the absence of unit-equations ($E_1=0$). 
$E_1(T)$ is a stochastic variable. However from the 
decoupling between fast and slow time scales sketched in Figure 
\ref{deterstoch} the probability distribution of $E_1(T)$ 
depends only on the slow time scale $t$. Let us call $\mu (E_1;t)$ 
this probability. Multiplying (\ref{probanocttot}) over the times
$T=0$ to $T=N-1$ we deduce the probability that DPLL has successfully 
found a solution without ever backtracking, 
\begin{equation} \label{rhosuc}
\rho _{success} = \exp \left( - \int _0^1 \frac{dt}{2(1-t)} 
\sum _{E_1\ge 1} \mu (E_1;t)\; (E_1-1) \right)
\end{equation}
in the large $N$ limit.

\begin{figure}
\begin{center}
\includegraphics[width=200pt,angle=0]{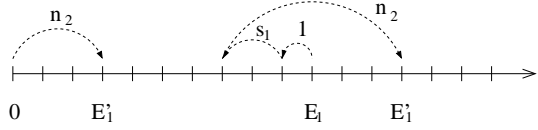}
\caption{Evolution of the number $E_1$ of 1-equations as one more variable
is assigned. $n_2$ denotes the number of 2-equations reduced to
1-equations, $s_1$ the number of 1-equations satisfied. If $E_1\ge 1$ a
variable is fixed through unit-propagation: $E_1$ decreases by one 
plus $s_1$, and increases by
$n_2$. In the absence of unit-equation ($E_1=0$) the number of
1-equations after the assignment is simply $E_1'=n_2$.  }
\label{rw1clause}
\end{center}
\end{figure}

We are left with the calculation of $\mu$\cite{Fr96}. 
Figure \ref{rw1clause} sketches the stochastic evolution of the number
$E_1$ during one step.  The number of 1-equations produced
from 2-equations, $n_2$, is a Poisson variable
with average value, from (\ref{sole23}),
\begin{equation}\label{valued}
d(t) = \frac {2 \, e_2 (t)}{1-t} = 6\,\alpha_0\, t(1-t)\,
\end{equation}
when $N\to \infty$. The number of 
satisfied 1-equations, $s_1$, is negligible as long as $E_1$ remains
bounded. The probability that the number of 1-equations
goes from $E_1$ to $E_1'$ when $T\to T+1$ defines the entry of the
transition matrix
\begin{equation}\label{matrixmm}
M (E'_1,E_1;t) = \sum _{n_2 \ge 0} e^{-d(t)} \frac{d(t)^{n_2}}{n_2!}
\delta _{E'_1-(E_1+n_2-\delta_{E_1})}\ .
\end{equation}
from which a master equation for the probability of
$E_1$ at time $T$ may be written. 
On time scales $1\ll \Delta T\ll N$ this master equation
converges to the equilibrium distribution $\mu$  \cite{Fr96,Co01},
conveniently expressed in terms of the generating function
\begin{equation} \label{defgtstat}
G( x;t ) = \sum _{E_1\ge 0} \mu(E_1;t)\; x^{E_1} = 
\frac{(1-d(t))(x-1)}{x\; e^{d(t) \; (1-x)} -1} \;
 \ . 
\end{equation}
The above is a sensible result for $d(t)\le 1$ but does not make sense when 
$d(t)>1$ since a probability cannot be negative! The reason is that we
have derived (\ref{defgtstat}) under the implicit condition that no
contradiction was encountered. This assumption cannot hold when the
average rate of 1-equation production, $d(t)$, is larger that one, the
rate at which 1-equations are satisfed by unit-propagation.
From (\ref{valued}) we see, when $\alpha > \alpha _E=\frac 23$, the
trajectory would cross the
\begin{equation}\label{defcon}
\alpha_D(p) = \frac {1} {2(1-p)}
\end{equation}
on which $d=1$ for some time $t_D<1$. A contradiction is very
likely to emerge before the crossing.

When $\alpha <\alpha _E$ $d$ remains smaller than unity at any time. 
In this regime the probability of
success reads, using (\ref{rhosuc}) and (\ref{defgtstat}),
\begin{equation} \label{nobtsuc}
\rho_{success} = \exp \left(\frac {3\alpha}{4} - \frac 12 
\sqrt{\frac{3\alpha}{2-3\alpha}}\; \tanh^{-1}
\bigg[ \sqrt{\frac{3\alpha}{2-3\alpha}}\bigg] \right)\ .
\end{equation}  
$\rho_{success}$ is a decreasing
function of the ratio $\alpha$, down from unity for $\alpha =0$ to zero for
$\alpha =\alpha_{E}$.
The present analysis of the UC heuristic can be easily transposed to the GUC
heuristic. Details are not given here but can be found in 
\cite{Ac02,Co01}. The result is an expression
for $\rho_{success}$ larger than its UC counterpart (\ref{nobtsuc}),
and vanishing in $\alpha _E\simeq 0.7507$. Interestingly the way 
$\rho_{success}$ vanishes when $\alpha$ reaches $\alpha_E$,
\begin{equation}
- \ln \rho_{success} (\alpha_E - \epsilon) \sim \epsilon ^{-\frac 12}
\qquad (\epsilon \to 0^+)
\end{equation}
is the same for both heuristics. This similarity extends to a whole
class of heuristics which can be described by the flow of equation
densities only and based on unit-propagation \cite{De04}. The
probability that DPLL finds a solution without backtracking
to a 3-XORSAT instance of size $N$ satisfies finite-size scaling at the 
dynamical critical point,
\begin{equation}\label{ffss}
- \ln \rho_{success} (\alpha_E - \epsilon, N) \sim N^{\frac 16} \;
\Phi\big( \epsilon \, N^{\frac 13} \big) \ ,
\end{equation}
where the scaling function $\Phi$ is independent of the heuristics
and can be calculated exactly \cite{De04}. The exponent
characterizing the width of the critical region is the one associated
to percolation in random graphs (\ref{valpsi}). A consequence of
(\ref{ffss}) is that, right at $\alpha_E$, $\rho_{success} \sim \exp 
(- Cst\times N^{\frac 16})$ decreases as a stretched exponential of
the size. The value of the exponent, and its robustness against the
splitting heuristics can be understood from the following
argument  \cite{De04}.

Let us represent 1- and 2- equations by a graph $G$ over the set
of $N-T$ vertices (one for each variable $x_i$) with $E_1$ marked 
vertices (one for each unit-equation $x_i=0,1$), and $E_2$
signed edges ($x_i + x_j=0,1$), see Section \ref{secrg}.
$d$ is simply the average degree of vertices in $G$.
Unit-propagation corresponds to removing
a marked vertex (and its attached edges), after having marked its
neighbours; the process is iterated until the connected component is
entirely removed (no vertex is marked). Meanwhile, new edges 
have been created from the reduction of 3-equations into 2-equations.
Then a vertex is picked up according to the heuristic and marked,
and unit-propagation resumes. The
success/failure transition coincides with the percolation
transition on $G$:  $d=1$ as expected.
From random graph theory \cite{Bo89} the
percolation critical window is of width 
$|d -1| \sim N^{-1/3}$. As $d$ is
proportional to the ratio $\alpha _0$ (\ref{valued})
we find back $\psi= \frac 13$.
The time spent by resolution trajectories in the critical
window is $\Delta t \sim \sqrt {|d -1|} \sim N^{-1/6}$,
corresponding to $\Delta T = N \, \Delta t \sim N^{5/6}$ eliminated
variables. As the largest components have size $S\sim N^{2/3}$ the
number of such components eliminated is $C=\Delta T/S\sim N^{1/6}$.
What is the probability $q$ that a large component is removed without
encountering a contradiction? During the removal of the component the
number of marked vertices `freely' diffuses, and reaches $E_1\sim \sqrt
{S}\sim N^{1/3}$. The probability that no contradiction occurs
is, from (\ref{probanocttot}), $q\sim (1-\frac {Cst}N)^{E_1\times S}$,
a finite quantity. Thus $\rho_{success} \sim q^C \sim
\exp(-N^{1/6})$. The presence of numerous, smaller components does 
not affect this scaling.

\subsection{Exponential phase:  massive backtracking}

For ratios $\alpha _0>\alpha _E$ DPLL is very likely to find a
contradiction. Backtracking enters into play, and is responsible for
the drastic slowing down of the algorithm (Figure \ref{dpllxorsat}).

\begin{figure}
\begin{center}
\includegraphics[width=130pt,angle=-90]{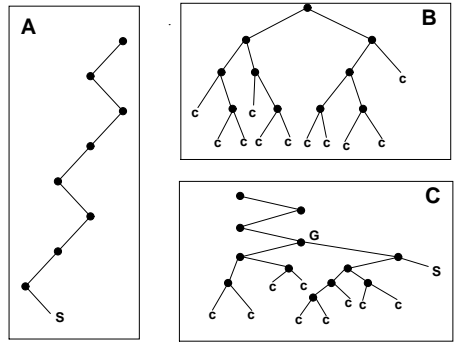}
\caption{Search trees in three regimes of Section \ref{secdpll}: 
{\bf A.} linear, satisfiable ($\alpha < \alpha_E$); {\bf  B.} 
exponential, satisfiable ($\alpha_E<\alpha<\alpha_c$);
{\bf C.} exponential, unsatisfiable ($\alpha > \alpha _c$). Leaves
are marked with S (solutions) or C (contradictions). G is the highest
node to which DPLL backtracks, see Figure \ref{diag2+p}.}
\label{fig-tree}
\end{center}
\end{figure}

The history of the search process can be represented by a search tree,
where the nodes represent the variables assigned, and the descending
edges their values (Figure \ref{fig-tree}). 
The leaves of the tree correspond to solutions (S), or
to contradictions (C). The analysis of the $\alpha<\alpha_E$ regime
leads us to the conclusion that search trees look like
Figure \ref{fig-tree}A at small ratios\footnote{A small amount of
  backtracking may be necessary to find the solution since
  $\rho_{success}<1$ \cite{Fr96}, but the overall picture of a single
  branch is not qualitatively affected.}. 
Consider now the case of unsatisfiable formulas 
($\alpha_0> \alpha_c$) where all leaves carry contradictions after
DPLL halts (Figure \ref{fig-tree}C). 
DPLL builds the tree in a sequential manner, adding nodes
and edges one after the other, and completing branches through
backtracking steps. We can think of the same search tree built in a 
parallel way\cite{Co01}. At time (depth $T$) our tree is 
composed of $L(T)\le 2^T$ branches, each carrying a partial assignment 
over $T$
variables. Step $T$ consists in assigning one more variable to each
branch, according to DPLL rules, that is, through unit-propagation or
split. Possible consequences are: emergence of a contradiction and end
of the branch, simplification of the attached formulas and the branch keeps
growing. 

The number of branches $L(T)$ is a stochastic variable. Its average
value can be calculated as follows \cite{Mon05}. Let us define the
average number $L(\vec E;T)$ of branches with equation populations $\vec
E=(E_1,E_2,E_3)$ at depth $T$. Initially $L(\vec E;0)=1$ for
$\vec E=(0,0,\alpha_0 N)$, 0 otherwise.
Call $M(\vec E',\vec E;T)$ the average
number of branches with population $\vec E'$ generated from a branch with
population $\vec E$ once the $T^{th}$ variable is assigned. Transition 
matrix $M$ is an extension of (\ref{matrixmm}) 
to the whole population vector $\vec E$ and not only $E_1$.
We have $0\le M\le 2$, the extreme values corresponding to a
contradiction and to a split respectively. We claim that 
\begin{equation}\label{evol5}
L(\vec E';T+1) = \sum _{\vec E} M(\vec E',\vec E;T) \; L(\vec E;T) \ .
\end{equation}
Evolution equation (\ref{evol5}) is somewhat suspicious since it looks like
the approximation (\ref{statq}) we have done in the analysis of
RWSAT. Yet a major difference exists which makes (\ref{evol5})
exact \cite{Mon05}. Drawing randomly many times the same instance, as
we are doing, is in principle forbidden but not along {\em one} branch 
for the very reason the analysis of Section \ref{secuc} was
correct. Actually what we have done in Section \ref{secuc} is to draw
randomly at time $T$ the equations containing the $T^{th}$
variable. But this is correct since those equations are immediately
simplified into shorter equations and their remaining content remains
unknown \cite{Ka02}. The situation seems more complicated in the case
of the whole tree since the same equation can appear at different
depth along distinct branches. Indeed the number of branches produced
from two distinct branches after assignment of one variable are correlated
variables. But thanks to the linearity of expectation those
correlations do not matter and (\ref{evol5}) is correct.

Transition matrix $M$ can be explicitely written down. It is more
convenient to write (\ref{evol5}) for the generating function of the
number of branches, $B(\vec x;T) = \sum _{\vec E} L(\vec E;T)
x_1^{E_1}\,x_2^{E_2}\,x_3^{E_3}$, with the result
\begin{equation} \label{evol6}
B(\vec x;T+1) = \frac 1{f_1} B(\vec f;T) + (2-\frac 1{f_1})
B(0,f_2,f_3;T) - 2 B(\vec 0;T)\ ,
\end{equation}
where $\vec f$ is the vector with components
\begin{equation}
f_1= x_1 + \frac{\frac 12-x_1}{N-T}, f_2 = x_2 +  \frac{2(
  x_2-x_1)}{N-T}, f_3 = x_3 +  \frac{3( x_2-x_3)}{N-T} \ .
\end{equation}
The three terms on the r.h.s. of (\ref{evol6}) correspond, from left
to right: unit-propagation (the branch keeps growing), variable
splitting (2 branches are created from the previous one), branches
carrying empty instances (satisfied instance).
Equation (\ref{evol6}) together with the initial condition $B(\vec
x;0)= x_3^{\alpha _0N}$ completely defines the average dynamics of the 
search tree. We sketch the main steps of its resolution below\cite{Co01}:
\begin{enumerate}
\item To count the  number of branches irrespectively of the number of 
unit-equations we should consider the value $x_1=1$. However, as long 
as branches grow the number $E_1$ of unit-equations cannot be large, 
and remains bounded. We can therefore choose $x_1=\frac 12$ which
simplifies (\ref{evol6}) without affecting the large size scaling of
$L$ and $B$. This technical trick is reminiscent of Knuth's kernel
method \cite{Kn66}.
\item For large $N$ it is reasonable to expect that the number of
  branches grows exponentially with the depth, or, equivalently,
\begin{equation}
\sum _{E_1} L(E_1,E_2,E_3;T) \sim e ^{N \; \lambda (e_2,e_2;t) + o(N)}
\end{equation}
where $e_2,e_3$ are the densities of equations as usual. From point 1
the Legendre transform of $\lambda$
\begin{equation}\label{degpp}
\gamma (x_2,x_3;t) = \max _{e_2,e_3} \big[ \lambda (e_2,e_2;t) + 
e_2 \ln x_2 + e_3 \ln x_3 \big]
\end{equation}
fulfills the partial differential equation (PDE)
\begin{equation}
\frac{\partial \gamma} {\partial t}  = \ln 2 + \frac{1-2x_2}{1-t} 
\frac{\partial \gamma}{\partial x_2} +  \frac{3(x_2-x_3)}{1-t} 
\frac{\partial \gamma}{\partial x_3} \ .
\end{equation}
with the initial condition $\gamma (x_2,x_3;t) = \alpha _0 \ln x_3$.
\item The first order PDE can be solved exactly with the
  characteristic method. The output, after Legendre inversion
 through (\ref{degpp}), is the entropy $\lambda (e_2,e_3;t)$ 
of branches at reduced depth $t$.  Let us call $\lambda ^*(t)$ the
  maximum value of $\lambda$ over the equation densities for a fixed
  fraction $t$ of assigned variables.
\item $\lambda ^*(t)$ is a function growing from $\lambda ^*=0$ at
$t=0$, reaching a maximum value $\lambda ^*_M$ in $t_M$, and
decreasing for larger times $t\le 1$. $t_M$ is the depth in the tree
of Figure \ref{fig-tree}C where most contradictions are found; the
number of C leaves is, to exponential order, $e^{N\lambda^*_M}$. We
conclude that the size of the tree we were looking for is
\begin{equation}  
\tau _{split} = \lambda^*_M \ ,
\end{equation}
compare with (\ref{mesure}). For large $\alpha \gg \alpha_c$ one finds
$\tau_{split}\sim \ln 2 /(6\alpha)$ in agreement with \cite{beame}. 
The calculation can be extended to highers values of $K$.
\end{enumerate}

The above calculation holds for the unsatisfiable, exponential phase.
How can we understand the satisfiable but exponential regime
$\alpha_E<\alpha_0<\alpha_c$? The resolution trajectory crosses the
Sat/Unsat critical line at some point G shown in Figure
\ref{diag2+p}. Immediately after $G$ the instance left by DPLL is
unsatisfiable. A subtree with all its leaves carrying contradictions
will develop below G (Figure \ref{fig-tree}B). The size $\tau_{split}
^G$ of this
subtree can be easily calculated from the above theory. The only 
change is the initial condition over $\gamma$: 
$\gamma (x_2,x_3;0) = \alpha_G (p_G \ln
x_3 + (1-p_G) \ln x_2)$ where $(p_G,\alpha_G)$ are the coordinates of
G which can be calculated from $\alpha_0$ and the knowledge of the
critical Sat/Unsat line. Once this subtree has been built DPLL
backtracks to G, flips the attached variable and will finally end up
with a solution. Hence the (log of the)  number of splits necessary will be
typically equal to $\tau_{split} = (1-t_G)\, \tau_{split}^G$ \cite{Co01}. 


\section{Conclusions}\label{secconclusion}

Previous Sections have allowed us to illustrate rather general
techniques and ideas to deal with random systems. It does not come as
a surprise that other problems than XORSAT e.g. the satisfaction of
Boolean constraints, graph coloring, the covering of vertices, ...  
have been successfully studied with these tools. Many of those
problems, when given 
an input distribution based on random graphs, actually share a lot 
of common features with XORSAT. The reader 
is referred to 
\cite{Mo97,Bi00,Me02,Me03b,Se06,Ac05} (satisfiability), 
\cite{Mu02,Se06} (coloring), 
\cite{We01b,We01} (vertex cover), ... 
for entry points to the literature. Let us also mention that 
many other interesting optimization problems, not directly related to
random graphs,  have been studied with the techniques of Sections 4
and 5, and the results sometimes rigorously proven 
e.g. matching \cite{Or85,Me87b,Al01}, 
traveling salesman \cite{Me86}, 
number partitioning \cite{Me98,Me01}, 
graph partitioning \cite{Fu85}, ...
Finally, from a historical point of view, one should not forget
that statistical mechanics tools have found numerous and beautiful
applications in the study of the learning and storage
properties of neural networks\cite{Am89,Va01}, all the more 
so the random satisfiability problem can be recast as an Ising
perceptron problem \cite{Kr89}.

The study of random optimization problems is obviously interesting
from a probabilistic point of view. As far as computer science is
concerned they can be seen as useful benchmarks for testing and
improving resolution procedures. A successful example is the
traduction of the cavity equations of Section 5 into an algorithm for
solving given instances of the satisfiability problem
\cite{Me02}. This algorithm, called Survey Propagation, extends to 
the clustered phase the Belief Propagation procedure of wide-spread use in 
statistical inference, and is a very efficient procedure
to find solutions to 
3-Satisfiability slightly below threshold. Another application of
statistical physics ideas is the conception of new heuristics for DPLL
capable of proving the unsatisfiability of formulas with 700 hundreds
variables at threshold \cite{De03}. 

Despite those successes important question remain open. First is there
a relationship between clustering and hardness of resolution? 
This question is reminiscent of a very general issue in statistical physics,
namely the relationship between dynamical and static properties of 
disordered or glassy systems \cite{Cu93}. 
The onset of clustering, or more
precisely of strong correlations between variables over the space of solutions
drastically worsens the performances of 
sampling algorithms e.g. Monte Carlo procedures \cite{Mo05,Se06}. However, in 
practical applications, one
looks for a solution rather than for the sampling of 
the solution space... From this point of view knowing whether 
solutions are clustered or not
 does not seem to be of crucial relevance. Actually 
a local and polynomial search strategy capable
of finding solutions well above the clustering threshold has been
explicitely found for various optimizations problems \cite{Jo07}.

Another open question is what happens at large
$K$, that is, when constraints involve more and more variables. 
The performances of all known algorithms, be they local search
procedures or DPLL solvers, seem to deteriorate. Worst-case bound 
indicate that the large $K$ case is very difficult \cite{Im99}. 
From statistical
mechanics point of view problems look like more and more the random
energy model \cite{Me87} as $K$ increases, but can we beat the
worst-case bounds on average? Finally let us mention a recent work by
Feige \cite{Fe02} which, for the first time, showed that the
complexity of solving random SAT (or XORSAT) model  had a fundamental
interest in worst-case approximation theory. Consider 3-SAT
instances with ratio $\alpha \gg \alpha_c$. Most of them have GS
energy close to $\alpha N/2$, but a very tiny fraction of those
instances have energy smaller than, say, $\epsilon N$ where $\epsilon\ll
\alpha$ is fixed. Is there a polynomial algorithm capable of
recognizing all such atypical formulas from the vast majority of
typical instances? Insights from statistical physics suggest that,
the answer is positive for SAT (if we want most satisfiable instances
to be detected and not all of them) while XORSAT seems to be much
harder\cite{Al07}! Actually, to the knowledge of the author, no local
search algorithm (based on random walk, variable assigment, Monte
Carlo, message-passing, cooling procedure, ...) is efficient for 
solving XORSAT. This makes the
study of this problem even more valuable from a computer science point
of view.


\appendix

\section{A primer on large deviations}\label{applargedev}

Large deviation theory is the field of probability which deals with
very unlikely events \cite{De93}. 
You are given a fair (unbiased) coin and
toss it $N$ times. The number $H$ of head draws has  probability
\begin{equation} \label{ld1}
p_N(H) = \frac 1{2^N} {N \choose H} \ .
\end{equation}
When $N$ gets large $H$ is highly concentrated around $H^*=N/2$
with small relative fluctuations of the order of $O(\sqrt N)$. 
Yet we can ask for the
probability of observing a fraction $h=H/N$ equal to say, 25\%, of
heads, far away from the likely value $h^*=50\%$. To calculate this
probability we use Stirling's asymptotic expression for the binomial
coefficient in (\ref{ld1}) to obtain
\begin{equation} \label{ld2}
p_N(H=h \, N) = e^{ - N \omega(h) +o(N)} \ ,
\end{equation}
where
\begin{equation}
\omega (h) = \ln 2 + h \ln h + (1-h)\ln (1-h)\ 
\end{equation}
is called rate function. The
meaning of (\ref{ld2}) is that events with value of $h\ne h^*$ are
exponentially rare in $N$, and $\omega(h)$ give the decay (rate) exponent.
The answer to our question is $e^{-N
\omega(.25)} \sim e^{-0.13 \,N}$ when $N$ is large. Some comments are:
\begin{itemize}
\item  $\omega(h)$ is strictly positive, except in $h = h^*=\frac 12$
  where it vanishes. This is the only value for the fraction of head draws
with non exponentially small--in--$N$ probability.
\item Let $h=h^* + \delta h$ where $\delta h$ is small. Using
  $\omega (h^*) = \omega '(h^*) =0$ we have
\begin{equation}
P_N \big( H= (h^* + \delta h) N \big) = \exp \big[ -N\, \frac 12
  \omega ''(h^*) \,(\delta h)^2 + \ldots \big] \ ,
\end{equation}
that is, $\delta h$ is Gaussianly distributed with zero mean and
variance $(N\omega''(h^*))^{-1}=(4N)^{-1}$. Hence central limit
theorem is found back from the parabolic behaviour of the
rate function around its minimum\footnote{Non standard behaviour
  e.g. fluctuations of the order of $N^{\nu}$ with $\nu\ne \frac 12$
as found in Levy flights correspond to non-analyticies of $\omega$ 
in $h^*$ or the vanishing  of the second derivative.}.
\item $\omega$ is here a convex function of its argument. This
  property is true rate functions describing independent
  events. Indeed, suppose we have $H$ positive (according to some
  criterion e.g. being a head for a coin) events among a set of $N$
  events, 
then another set of $N'$ events among which $H'$ are positive. If the 
two sets are uncorrelated 
\begin{equation}
p_{N+N'} (H+H') \ge p_N (H) \times p_{N'} (H')
\end{equation}
since the same total number $H+H'$ of positive events could be
observed in another
combination of $N+N'$ events. Taking the logarithm 
and defining $h=H/N$, $h'=H'/N$, $u=N/(N+N')$ we obtain
\begin{equation}
\omega (u\, h+ (1-u)\, h') \le u\, \omega(h) + (1-u)\, \omega(h') \ ,
\end{equation}
for any $u\in [0;1]$. Hence the representative curve of $\omega$ lies
below the chord joining any two points on this curve, and $\omega$ is
convex. Non-convex rate functions are found in presence
of strong correlations\footnote{Consider the following experiment. 
You are given three
  coins: the first one is fair (coin A), the second and third coins,
  respectively denoted by B and C,  are biased and
  give head with probabilities, respectively, $\frac 14$ and $\frac
  34$. First draw coin A once. If the outcome is head pick up coin B,
  otherwise pick up coin C. Then draw your coin $N$ times. What is the
  rate function associated to the fraction $h$ of heads?}.
\end{itemize}

\section{Inequalities of first and second moments}\label{appmoment}

Let ${\cal N}$ be a random variable taking values on the positive integers, and
call $p_{\cal N}$ its probability. We denote by $\langle{\cal N}\rangle$ and 
$\langle {\cal N} ^2\rangle$ the first and second moments of 
${\cal N}$ (assumed to be finite), and write 
\begin{equation}
p({\cal N}\ge 1) =\sum _{{\cal N}=1,2,3, \ldots} p_{\cal N} = 1 -p_0
\end{equation}
the probability that ${\cal N}$ is not equal to zero.
Our aim is to show the inequalities
\begin{equation} \label{ineq}
\frac{\langle {\cal N}\rangle^2}{\langle {\cal N}^2\rangle} \le  
p({\cal N}\ge 1) \le {\langle {\cal N}\rangle} \ .
\end{equation}
The right inequality, call 'first moment inequality', is straightforward:
\begin{equation}
{\langle {\cal N}\rangle} = \sum _{\cal N} {\cal N}\; p_{\cal N} = 
\sum _{{\cal N}\ge 1} {\cal N} \; p_{\cal N} \ge
\sum _{{\cal N}\ge 1}  p_{\cal N} =  p({\cal N}\ge 1) .
\end{equation}
Consider now the linear space made of vectors 
${\bf v}=(v_0,v_1,v_2,\ldots\}$ whose components are labelled by positive 
integers, with the scalar product
\begin{equation}
{\bf v} \cdot {\bf v}' = \sum _{\cal N} p_{\cal N}\; v_{\cal N}\; 
v_{\cal N}' \ .
\end{equation}
Choose now $v_{\cal N}={\cal N}$, and $v'_0=0,v'_{\cal N}=1$ for 
${\cal N}\ge 1$. Then
\begin{equation}
{\bf v} \cdot {\bf v} = \langle {\cal N}^2\rangle \ , \ 
{\bf v} \cdot {\bf v}' = \langle {\cal N}\rangle \ , \
{\bf v}' \cdot {\bf v}' = p({\cal N}\ge 1) \ .
\end{equation}
The left inequality in (\ref{ineq}) is simply the Cauchy-Schwarz inequality 
for ${\bf v},{\bf v}'$: $({\bf v} \cdot {\bf v}')^2 \le  ({\bf v} \cdot {\bf v} ) \times ({\bf v}' \cdot {\bf v} ')$. 

\section{Corrections to the saddle-point calculation of $\langle 
{\cal N}^2\rangle$} \label{appfluctu}

In this Appendix we show that $\langle {\cal N}^2\rangle$ is
asymptotically equivalent to $\langle {\cal N}\rangle^2$, where ${\cal N}$ is the number of solutions of a 3-XORSAT formula with ratio $\alpha 
< \alpha_2 \simeq 0.889$. This
requires to take care of the finite-size corrections around the
saddle-point calculations of Section \ref{secld2}. Let $Z=(z_1,z_2,
\ldots, z_N)$ denotes a configuration of variables at distance $d$
from the zero configuration {\em i.e.} $dN$ variables $z_i$ are equal
to 1, the other  $(1-d)N$ variables are null. 
Let $q(d,N)$ be the probability that $Z$ satisifies 
the equation $z_i+z_j+z_k=0$ where $(i,j,k)$ is a random triplet of
distinct integers (unbiased distribution):
\begin{eqnarray}
q(d,N)&=& \frac 1{{N\choose 3}} \left[ {(1-d)N \choose 3} + (1-d)N \;
{dN \choose 2} \right] \\
&=& q(d) \left( 1+ \frac{h(d)}N\right) + 
\ldots \quad \mbox{where} \quad h(d) =\frac{
6 d  (2d-1)}{3 d^2+(1-d)^2}\ ,\nonumber 
\end{eqnarray}
and $q(d)$ is defined in (\ref{qk}) with $K=3$. Terms of the order of $N^{-2}$ 
have been discarded. 

Using formula (\ref{n2p}) with $q(d)$ substituted with $q(d,N)$ 
and the Stirling formula for the asymptotic behaviour of
combinatorial coefficients we have
\begin{eqnarray} \label{truc1}
\langle {\cal N}^2\rangle \sim \sum _{d=0,\frac 1N, \frac 2N, \ldots} 
\frac{\sqrt{2\pi N}}{\sqrt{2\pi Nd}\sqrt{2\pi N(1-d)}}\;
e^{N\, A(d,\alpha) + \alpha\, h(d)}
\end{eqnarray}
where $A(d,\alpha)$ is defined in (\ref{upperom2}), and $\sim$
indicates a true asymptotic equivalence (no multiplicative factor
omitted). The r.h.s. of (\ref{truc1}) is the Riemann sum associated 
to the integral
\begin{eqnarray} \label{truc2}
\langle {\cal N}^2\rangle \sim \int _0^1 \frac{N \, dd}
{\sqrt{2\pi Nd(1-d)}}\;
e^{N\, A(d,\alpha) + \alpha\, h(d)} \ .
\end{eqnarray}
We now estimate the integral through the saddle-point method.
For $\alpha < \alpha _2 \simeq 0.889$ the dominant contribution to the integral comes from the vicinity of
$d^*=\frac 12$. There are quadratic fluctuations 
around this saddle-point, with a variance equal to $N$ times the 
inverse of (the modulus of) 
the second derivative $A_{dd}$ of $A$ with respect to $d$.
Carrying out the Gaussian integral over those fluctuations we obtain
\begin{equation} \label{truc3}
\langle {\cal N}^2\rangle \sim  \frac{N \;
e^{N\, A(d^*,\alpha) + \alpha\, h(d^*)} }{\sqrt{2\pi N d^*(1-d^*)}}\;
\sqrt{\frac{2\pi}{N\, |A_{dd}(d^*,\alpha)|}}
\sim  \langle {\cal N}\rangle ^2
\end{equation}
since $h(d^*)=0$, $A_{dd}(d^*,\alpha)=-4$. Therefore, from the
second moment inequality, $P_{SAT}\to 1$ when $N\to\infty$ at ratios
smaller than $\alpha_2$.

%
%

\end{document}